\documentclass[aps,prb,twocolumn,amsmath,showpacs,amssymb]{revtex4}
\usepackage{amsmath}
\usepackage{graphicx}
\usepackage{amssymb}

\newcommand{\pdag}{{\phantom{\dagger}}}
\usepackage{dsfont}

\newcommand{\be}{\begin{equation}}
\newcommand{\ee}{\end{equation}}

\begin{document}

\title{Helical nuclear spin order in a strip of stripes in the Quantum Hall regime}

\author{Tobias Meng}
\affiliation{Department of Physics, University of Basel, Klingelbergstrasse 82, CH-4056 Basel, Switzerland}
\author{Peter Stano}
\affiliation{Center for Emergent Matter Science, RIKEN, Wako, Saitama 351-0198, Japan}
\affiliation{Institute of Physics, Slovak Academy of Sciences, 845 11 Bratislava, Slovakia}
\author{Jelena Klinovaja}
\affiliation{Department of Physics, Harvard University, Cambridge, Massachusetts 02138, USA}
\author{Daniel Loss}
\affiliation{Department of Physics, University of Basel, Klingelbergstrasse 82, CH-4056 Basel, Switzerland}




\begin{abstract}
We investigate nuclear spin effects in a two-dimensional electron gas  in the quantum Hall regime modeled by a weakly coupled array of interacting quantum wires. We show that the presence of hyperfine interaction between electron and nuclear spins in such wires can induce a phase transition, ordering electrons and nuclear spins into a helix in each wire. Electron-electron interaction effects, pronounced within the one-dimensional stripes, boost the transition temperature up to tens to hundreds of millikelvins in GaAs. We predict specific experimental signatures of the existence of nuclear spin order, for instance for the resistivity of the system at transitions between different quantum Hall plateaus.
\end{abstract}
%
%

\maketitle

\section{Introduction}
In contrast to their higher dimensional analogs, one-dimensional electron systems do not, in general, give rise to Fermi liquid phases. Rather than electronic quasiparticles, the low energy excitations are collective density waves (bosons), and the system can be described as a Luttinger liquid [\onlinecite{giamarchi_book}]. Among the hallmarks of this state are the separation of spin and charge degrees of freedom, and the power law decay of correlations.

Because of the underlying Luttinger liquid character, arrays of coupled one-dimensional electron systems(``stripes'') provide a promising platform to study non-Fermi liquids in two dimensions. While the hopping between neighboring stripes tends to restore Fermi liquid physics, two-dimensional Luttinger liquid phases (the smectic metals or sliding Luttinger liquids) have been shown to survive for specific forms of the interaction between stripes [\onlinecite{emery_00,vishwanath_01,mukhopadhyay_01}]. At filling factors away from values corresponding to quantum Hall fillings, a strong magnetic field is expected to stabilize such an anisotropic metallic state [\onlinecite{sondhi_01}]. At quantum Hall fillings, the integer [\onlinecite{sondhi_01,kane_02}] and fractional [\onlinecite{klinovaja_strip_of_stripes_lattice,klinovaja_strip_of_stripes,kane_02,kane_arxiv}] quantum Hall effects can be obtained from a stripe model with interstripe single particle hopping, or more general interaction processes between stripes. This way, the whole hierarchy of quantum Hall effects can be obtained within a stripe model [\onlinecite{klinovaja_strip_of_stripes_lattice}]. There are further hints on stripes from numerics: Hartree-Fock calculations showed that transitions between quantum Hall plateaus are susceptible to the formation of a charge density wave [\onlinecite{koulakov_96,fogler_96,moessner_96}], which allows for the emergence of a smectic metal phase [\onlinecite{fradkin_99,mcdonald_00}]. Finally, the observation of anisotropic responses gives experimental support for the existence of the quantum Hall stripe phase [\onlinecite{lilly_99,du_99,shayegan_00,kukushkin_11}]. In addition, the anisotropy is an intrinsic property of many quasi-two-dimensional materials such as organic compounds, for example, Bechgaard salts, in which quantum Hall effect has been studied [\onlinecite{Lebed,Montambaux,Lebed_Gorkov,Yakovenko_PRB,Lee_PRB,Yakovenko_review,Stripe_PRL_exp}]. For the use of wire construction beyond quantum Hall effect we refer to Refs.~[\onlinecite{klinovaja_strip_of_stripes_lattice,Non-Abelian,PF1,PF2,Thomale,Yaroslav}]. 

To establish the existence of stripes for a real two-dimensional electron gas more firmly, observations of effects unique to one dimension are necessary. One intriguing consequence of Luttinger liquid physics is the helical ordering of nuclear spins [\onlinecite{braunecker_nuclear_order_prl_09,braunecker_nuclear_order_prb_09,meng_nuclear_order_double_wire}], for which there is recent experimental evidence from transport measurements in a single quantum wire [\onlinecite{scheller_13}]. In this scenario, the electrons within a wire mediate a Rudermann-Kittel-Kasuya-Yosida (RKKY) exchange interaction [\onlinecite{RKKY_R,RKKY_K,RKKY_Y}] for the nuclear spins. This exchange has a Luttinger liquid peak at $2\,k_F$, where $k_F$ is the Fermi momentum, and drives an ordering of the nuclear spins in the form of a helix. In the remainder, we argue that a similar helical ordering can take place in smectic metals, and in particular at the edges of a quantum Hall plateau. Based thereon, we discuss experimental consequences of the formation of the helical order. Their observation would be a signature of non-Fermi liquid physics in two dimensions, and would support scenarios advocating the formation of quantum Hall states out of coupled Luttinger liquids. However, we would like to emphasize that the model considered in this work can also be engineered as an array of coupled one-dimensional systems such as nanowires, carbon nanotubes [\onlinecite{McEuen_CNT,Klinovaja_CNT_RKKY,braunecker_nuclear_order_prl_09,Marcus_CNT_nuclear,CNT_Schonenberger}], and quantum wires [\onlinecite{yacoby_2_wires,yacoby_wires_Science,scheller_13}]. In all these systems RKKY interaction is significantly strong and can be measured via local spin susceptibility [\onlinecite{Stano_measure_RKKY}].

The paper is organized as follows. Our model is defined in Sec.~\ref{sec:model}, while we analyze the ordering of the nuclear spins in Sec.~\ref{sec:ordering}. After a general discussion of the associated ordering temperature in Sec.~\ref{sec:Tc}, gapless systems at quantum Hall plateau edges are analyzed in Sec.~\ref{sec:IIIC}. In Sec.~\ref{sec:IIIE}, we collect some experimental signatures of the formation of helical nuclear spin order. Sec.~\ref{sec:IIIB} comments on the absence of nuclear spin order on the gapped quantum Hall plateaus, while the transition between the gapped and gapless regimes is discussed in Sec.~\ref{sec:IIID}. We close with a discussion of the physics of the nuclear spin order in Sec.~\ref{sec:disc}.

\section{The model}\label{sec:model}
\subsection{Strip of stripes model of a two-dimensional electron gas}
With the motivation given in the introduction, we adopt a model of the quantum Hall effect (QHE) based on an array of coupled stripes (in further also referred to as wires). We assume them to be aligned along the $x$ axis, which is perpendicular [\onlinecite{klinovaja_strip_of_stripes_lattice,klinovaja_strip_of_stripes}] to the edges of the sample (a different stripe alignment, along the longitudinal direction of the sample, was considered in Refs.~[\onlinecite{kane_02,kane_arxiv}]). We consider spin unpolarized electrons, corresponding to even numerator filling factors $\nu$ (the latter is defined as the ratio of the total number of electrons in the system and the number of degenerate states in each Landau level) [\onlinecite{footnote_pol}]. We introduce $\Psi_{n, \sigma}^\dagger(x) $ as the creation operator for an electron at position $x$ in stripe $n$ with spin index $\sigma=\uparrow,\downarrow$ (corresponding to $\sigma = +1,-1$ if used as a number), where the arrows refer to the direction of the applied magnetic field ${\bf B}=\nabla \times {\bf A}$ along the $z$ axis (perpendicular to the sample). In the Landau gauge, where ${\bf A}=(0,B x,0)$, independent and non-interacting stripes are described by the Hamiltonian density [\onlinecite{footnote_dens}]
\begin{equation}
\mathcal{H}_x  = \sum_{n, \sigma} \Psi^\dagger_{n,  \sigma}(x) \Big[-\frac{\hbar^2 \partial_x^2}{2m_e} - \mu + \sigma \Delta_Z\Big]  \Psi_{n, \sigma}(x),
\label{eq:Hx}
\end{equation}
with the effective electron mass $m_e$, the chemical potential $\mu$, the Zeeman splitting $\Delta_Z = S g\mu_B B$, the $g$-factor $g$, the Bohr magneton $\mu_B$, and the electron spin modulus $S=1/2$. As a first step, the non-interacting spectrum is projected onto the right and left moving modes close to the Fermi points using $\Psi_{n,\sigma}(x) \approx e^{ixk_{F\sigma}}R_{n,\sigma}(x)+e^{-ixk_{F\sigma}}L_{n,\sigma}(x)$, where $k_{F\sigma} = \sqrt{2m_e(\mu -\sigma \Delta_Z)}/\hbar$ is the Fermi momentum of spin $\sigma$ electrons. We then add Coulomb interactions to Eq.~\eqref{eq:Hx}, and retain only the dominant intrastripe interaction processes with zero momentum transfer. These are associated with a matrix element $U$, such that the Hamiltonian density describing electron-electron repulsion reduces to the form
\begin{equation}
\mathcal{H}_{C}=\frac{U}{2} \sum_{n} \Big[\sum_{r,\sigma} r^\dagger_{n,\sigma}(x)  r_{n,\sigma}^\pdag(x) \Big]^2~,\label{eq:coulomb_ferm}
\end{equation}
where $r=R,L$ labels the chirality. In order to tackle this interacting problem, we linearize the kinetic energy in $\mathcal{H}_x$ (now containing also $\mathcal{H}_C$) around the Fermi points. Thereby, we neglect the Zeeman splitting by setting $\Delta_Z \to 0$. We have checked that for the experimentally relevant fields of up to a few Tesla, the RKKY exchange calculated below exhibits only negligible corrections due to the Zeeman splitting of spin up and spin down electrons (a more detailed comment is given in Appendix \ref{sec:append_ham}). Using standard bosonization techniques [\onlinecite{giamarchi_book}], we arrive at
\begin{equation}
 \mathcal{H}_x =\frac{\hbar}{2\pi}\sum_{n,\kappa=\rho,s}\frac{u_\kappa}{K_\kappa}(\partial_x \phi_{n,\kappa})^2+u_\kappa K_\kappa(\partial_x \theta_{n,\kappa})^2~,\label{eq:llham1}
\end{equation}
where the fields $\phi_{n,\rho}$ and $\phi_{n,s}$ are proportional to the integrated charge and spin densities in stripe $n$, respectively, while the fields $\theta_{n,\kappa}$ are canonically conjugate to $\phi_{n,\kappa}$. The effective velocities in the spin and charge sectors, $u_\kappa$, and the associated Luttinger parameters, $K_\kappa$, are assumed to be identical for all stripes.

In addition to the motion along the stripes, the electrons are allowed to hop between the stripes with a hopping amplitude $t_y$,
\begin{equation}
\mathcal{H}_{y} = \sum_{n,\sigma} t_{y} e^{i\varphi(x)}\Psi^\dagger_{n+1,\sigma}(x)  \Psi_{n,\sigma}(x)+ {\rm H.c.}~,
\label{1d_y}
\end{equation}
where the phase $\varphi (x)$ is generated by the uniform magnetic field  ${\bf B}$ perpendicular to the $(x,y)$-plane. In the Landau gauge, we obtain  
\begin{equation}
\varphi(x) = \frac{e}{\hbar c} B x a_{y},
\label{phase_define} 
\end{equation}
where $a_y$ denotes the distance between stripes. If the magnetic field is such that the phase $\varphi(x)$ results in a resonant backscattering at the Fermi level (between states with momenta $+k_F$ and $-k_F$), a Peierls gap develops in the system [\onlinecite{braunecker_peierls}]. The phase $\varphi(x)$ picked up by an electron tunneling between any two stripes can then be understood as a momentum kick between the right and left Fermi point. Since this momentum kick does not depend on the stripe index, the resonance condition is independent of the longitudinal coordinate $n$. If, on the contrary, the magnetic phase is off resonant, no Peierls gap develops.  By considering also umklapp scatterings, the resonant tunneling between stripes can explain both the integer and fractional quantum Hall conductance hierarchies [\onlinecite{klinovaja_strip_of_stripes_lattice,klinovaja_strip_of_stripes}].  

\subsection{Nuclear spins and RKKY interaction}
In GaAs, our material of choice, each atomic nucleus has a spin. The dominant electron-nuclear coupling is given by the Fermi contact hyperfine interaction 
\begin{equation}
H_{I} = \frac{A}{N_\perp} \sum_{n,l}  { {\bf I}}_{n,l} \cdot  \Big[ S a \sum_{\alpha,\beta}  \Psi^\dagger_{n,\alpha}(x_l)  \boldsymbol{\sigma}_{\alpha\beta} \Psi_{n,\beta}(x_l) \Big],
\label{eq:Hhyp}
\end{equation}
where the nuclear spin ${\bf I}_{n,l}$ is inside stripe $n$ at a position with longitudinal coordinate $x_l$ and transverse coordinate ${\bf r}_{\perp,l}$. The latter refers to the position within the stripe cross-section. We parametrize the cross-section area $C$ by the number of nuclear spins within, $N_\perp=Ca\rho_0$, typically $N_\perp\gg1$, introducing a length scale $a$ of the order of the lattice constant $a_0$. We use $a$ as a short-distance cutoff. The nuclear spins ${\bf I}_{n,l}$ have a volume density $\rho_0$ and  length (in units of $\hbar$) $I$. The term in the bracket in Eq.~\eqref{eq:Hhyp} is the operator of the electron spin ${\bf S}_{n}(x_l)$, with $\boldsymbol{\sigma}$, the vector of Pauli matrices with matrix elements indexed by $\alpha, \beta = \uparrow, \downarrow$. The material dependent hyperfine coupling is given by $A$. In GaAs $a_0=0.565$ nm, $\rho_0=8/a_0^3$, $I=3/2$, and $A=90$ $\mu$eV. The much weaker dipolar interactions between nuclear spins are neglected [\onlinecite{braunecker_nuclear_order_prl_09,braunecker_nuclear_order_prb_09,meng_nuclear_order_double_wire}].

We treat the hyperfine coupling $H_I$ as a perturbation to the rest of the electron Hamiltonian $\mathcal{H}_x+\mathcal{H}_y$. Using linear response theory, we consider each nuclear spin as an independent source of electron spin polarization, which couples to the other nuclear spins. As we discuss below, the dominant RKKY exchange occurs between nuclear spins within the same stripe, and aligns all nuclear spins within a given cross section ferromagnetically. The exchange can thus be calculated along the lines of Ref.~[\onlinecite{braunecker_nuclear_order_prb_09}], which yields an effective interaction 
\begin{equation}
H_R=\sum_{n,i, j} {\bf \tilde{I}}_{n,i} \cdot {\bf \tilde{I}}_{n,j} J_{i j},
\label{eq:RKKY}
\end{equation}
where ${\bf \tilde{I}}_{n,i} = \sum_{l\in i} {\bf I}_{n,l}$ is the sum of all nuclear spins within $i$-th transverse plane, defined as a volume $Ca$ centered at the longitudinal coordinate $x_i$. The interaction strength is parametrized by $J_{i j}\equiv J(x_i-x_j)=J_{ji}$, the static RKKY coupling. In the lowest order expansion in $A/E_F$, with $E_F$ being the electron Fermi energy, $J$ scales with $A^2/E_F$, since $H_I$ is both the source of perturbation (spin ${\bf \tilde{I}}_{n,i}$) and the source of the energy gain (of spin ${\bf \tilde{I}}_{n,j}$). The functional dependence of the RKKY coupling is strongly influenced by a resonant backscattering at the Fermi energy. This is a general feature of a one dimensional electronic system, and is present for Fermi liquid, Luttinger liquid [\onlinecite{braunecker_nuclear_order_prb_09,braunecker_nuclear_order_prl_09,RKKY_1D,RKKY_Nanotube}], and even gapped phases, like the superconducting one [\onlinecite{klinovaja_sc,braunecker_sc,franz_sc}], or, as we will see here, the QHE. The resonant enhancement is reflected as a narrow dip at $k=2k_F$ of $J_k$, the Fourier transform of $J(x)$ expressed in terms of the momentum variable $k$. As a consequence, the ground state of the system is a nuclear helimagnet, where the nuclear spin orientation is uniform within a given cross section but rotates as one moves along the stripe,
\begin{equation}
\langle {\bf \tilde{I}}_{n,i} \rangle = m \, R_{{\bf h}, 2k_F x_{i}} \cdot {\bf \tilde{I}}_{n,0}.
\label{eq:ground state}
\end{equation}
Here, the c-number vector ${\bf \tilde{I}}_{n,0}$ gives the nuclear spin orientation at the longitudinal coordinate $x_{i}=0$, and $R_{{\bf h},\alpha}$ is a 3 x 3 matrix of rotation of a vector around axis ${\bf h}$ (referred to as the helical axis; it may depend on $n$, but we omit this index to ease the notation) by the angle $\alpha$. The orientation of ${\bf h}$ and ${\bf \tilde{I}}_{n,0}$ is at the moment unspecified, except for the requirement that they are perpendicular to each other. The angular brackets denote the expectation value taken with the system density matrix. The nuclear magnetization $m$, normalized to one, represents the order parameter, with $m=1$ for a fully ordered helical ground state, and $m=0$ for no order. 

Once the nuclear order is established, it corresponds to a macroscopic magnetic field (Overhauser field). If the electrons were initially gapless, this field has an important back action on the electrons mediating the RKKY exchange. The state of other nuclear spins can thus not be neglected anymore in calculating the RKKY exchange between a given pair of nuclear spins. We therefore add the mean value of Eq.~\eqref{eq:Hhyp},
\begin{equation}
\langle \mathcal{H}_I \rangle = \sum_n \frac{m A S I}{\pi a} \,\cos\left(\sqrt{2}(\phi_{n,\rho}+ \theta_{n,s})\right)~, \label{eq:ov_ham_bos}
\end{equation} 
into the electron Hamiltonian, and recalculate the RKKY coupling. The brackets on the left hand side imply that we have replaced the nuclear spin operators by numbers according to Eq.~\eqref{eq:ground state}, and that we have used the bosonization prescription to replace the remaining fermionic electron operators to arrive at the right hand side in terms of bosonic fields. We refer to adding (not adding) the Overhauser field into the unperturbed electron Hamiltonian as taking (not taking) the electron-nuclear feedback into account. Evaluating the RKKY coupling in the presence of the Overhauser field goes beyond the linear response formalism, though we are still able to treat it analytically. The Overhauser field opens a partial gap in the electron system with profound effects. The RKKY coupling is enhanced, the more the stronger the electron-electron interactions are. This enhancement can renormalize the ordering temperature by orders of magnitude [\onlinecite{braunecker_nuclear_order_prb_09,braunecker_nuclear_order_prl_09}].

\section{Ordering of the nuclear spins}\label{sec:ordering}

\subsection{Critical temperature}
\label{sec:Tc}
We now discuss the critical temperature of the nuclear order within a single stripe (we state here only main results, and refer the reader to Appendix \ref{app:Tc} for details). The critical temperature $T_c$ is found by examining excitations above the ground state of the system. This ground state is given by Eq.~\eqref{eq:ground state}. Similar to a uniform ferromagnet, the relevant excitations are bosonic spin waves (magnons). Due to the resonant shape of the RKKY exchange, however, the energies are somewhat unusual here. Magnons split into two classes: short, and long wavelength ones. The distinction is defined by comparing the magnon wave vector $q$ with $q_w$, the width of the RKKY exchange dip in momentum space. [A more precise definition is given below Eq.~\eqref{eq:magnon dispersion}].

Consider first the short wavelength magnons. To a good approximation, their energies are wave vector independent, and depend only on the value of the RKKY exchange at its minimum,
\begin{equation}
\epsilon_{2k_F} \approx \epsilon_M(m) \equiv -2 m N_\perp I J_{2k_F}. 
\label{eq:epsilonM}
\end{equation}
Since each magnon diminishes the order by the same amount as a flip of a single spin 1/2, if energies of all magnons are described by Eq.~\eqref{eq:epsilonM}, the system is equivalent to non-interacting spins in an effective field $g \mu_B B_{\rm eff} = \epsilon_M (m)$. The magnetization is then given by a self-consistent equation
\begin{equation}
m = B_I \left( \frac{\epsilon(m) I}{k_BT} \right),
\label{eq:magnetization definition}
\end{equation}
with $B_I$ denoting the Brillouin function [\onlinecite{kittel_book}], and where $\epsilon(m)$ is defined in Eq.~\eqref{eq:MPK} below. For $T\to 0$, the order is established, $m\to 1$. We define the critical temperature as $T$ for which the magnetization, given as a solution to Eq.~\eqref{eq:magnetization definition}, drops to a value close to 1/2 [\onlinecite{footnote1}].

Besides magnon energies, the energy $\epsilon(m)$ in Eq.~\eqref{eq:magnetization definition} may include additional contributions. For an initially gapless stripe, that is away from the quantum Hall plateau, we have identified two contributions,
\begin{equation}
\epsilon(m) = \epsilon_M(m) + \epsilon_P(m) + \epsilon_K(m),
\label{eq:MPK}
\end{equation} 
which appear from the back action of the established nuclear order on the electron system. Namely, a finite helical order $m>0$ leads to resonant backscattering of electrons at the Fermi energy and opens a partial gap. This leads to, first, a Peierls-like energy gain of the electron system [\onlinecite{braunecker_peierls}]. Second, the electron spin develops a finite polarization, locally collinear with the helical orientation, $\langle {\bf S}_n(x_i) \rangle \propto \langle {\bf I}_{n,i} \rangle$. This in turn results in a finite Knight field acting on the nuclear spins. These two energies, per a single flipped nuclear spin, are denoted as $\epsilon_P$, and $\epsilon_K$, respectively. Finally, also the value of the RKKY exchange is affected by the feedback effect through the opening of the partial gap. In another words, $J$ might also depend on $m$. We comment on this below for specific cases. If the electron system is already gapped before the nuclear spin order is formed, that is for a system on the quantum Hall 
plateau, these back action effects are absent, and $\epsilon(m)=\epsilon_M(m)$.

We now turn to the long-wavelength magnons. These have linear spectrum $\epsilon_q \approx \hbar c q$, with the velocity $c$ given by the curvature of $J_k$ at its minimum. Their number is
\begin{equation}
N_L=2\sum_{q=q_1}^{q_w} \frac{1}{\exp(\epsilon_q/k_B T)-1},
\label{eq:Tc2}
\end{equation}
where $q_1=2\pi/L$ is the minimal wave vector in a wire of length $L$ and the factor 2 is to account for contribution from negative $q$'s. In an infinite system, such bosonic excitations would mean that the order cannot be established at any finite temperature, since the sum diverges for $L\to \infty$. For a finite wire, the sum is finite, and so is the critical temperature. This also follows from a generalized Mermin-Wagner theorem for RKKY systems [\onlinecite{loss_legget_11}]. We define the critical temperature from long-wavelength magnons by equating their number to the ground state magnetization, $3N_L=I N N_\perp$ (the factor $3$ counts the long-wavelength magnon branches; see App.~\ref{app:Tc1}).

Although the short and long-wavelength magnons diminish the order together, a good estimate for the critical temperature is given by the minimum of the two critical temperatures obtained from the long and short wavelength magnon separately, calculated from Eq.~\eqref{eq:magnetization definition} and Eq.~\eqref{eq:Tc2}.  We show in App.~\ref{app:Tc1} how these two cases follow as limits from a common general formula for the statistical sum, Eq.~\eqref{eq:Z}.

Let us be more specific for the case of a QHE strip of stripes. There we find that, first of all, the Peierls and Knight field energies in Eq.~\eqref{eq:MPK} are negligible compared the the RKKY energy, such that $\epsilon(m)\approx\epsilon_M(m)$. For a gapless stripe, the long-wavelength magnons contribution is furthermore negligible for any realistic wire length. Though we were not able to obtain an analytical expression for the RKKY exchange away from the minimum for a gapped stripe, our calculations suggest that the long-wavelength magnons (if present at all) have negligible effects also here. For both a gapped and a gapless phase, the critical temperature is therefore given by Eqs.~\eqref{eq:epsilonM}-\eqref{eq:magnetization definition}, which, using a Taylor expansion for small magnetizations $m$, yield [\onlinecite{footnote1}]
\be
k_B T_c = \frac{2}{3} N_\perp I^{2} J_{2k_F}(T_c).
\label{eq:Tc final}
\ee
 We now proceed to calculate $J_{2k_F}$ necessary for the evaluation of the critical temperature.

\subsection{Nuclear order in a gapless stripe}
\label{sec:IIIC}
\begin{figure}
\centering
\raisebox{3cm}{(a)}\quad\includegraphics[scale=0.4]{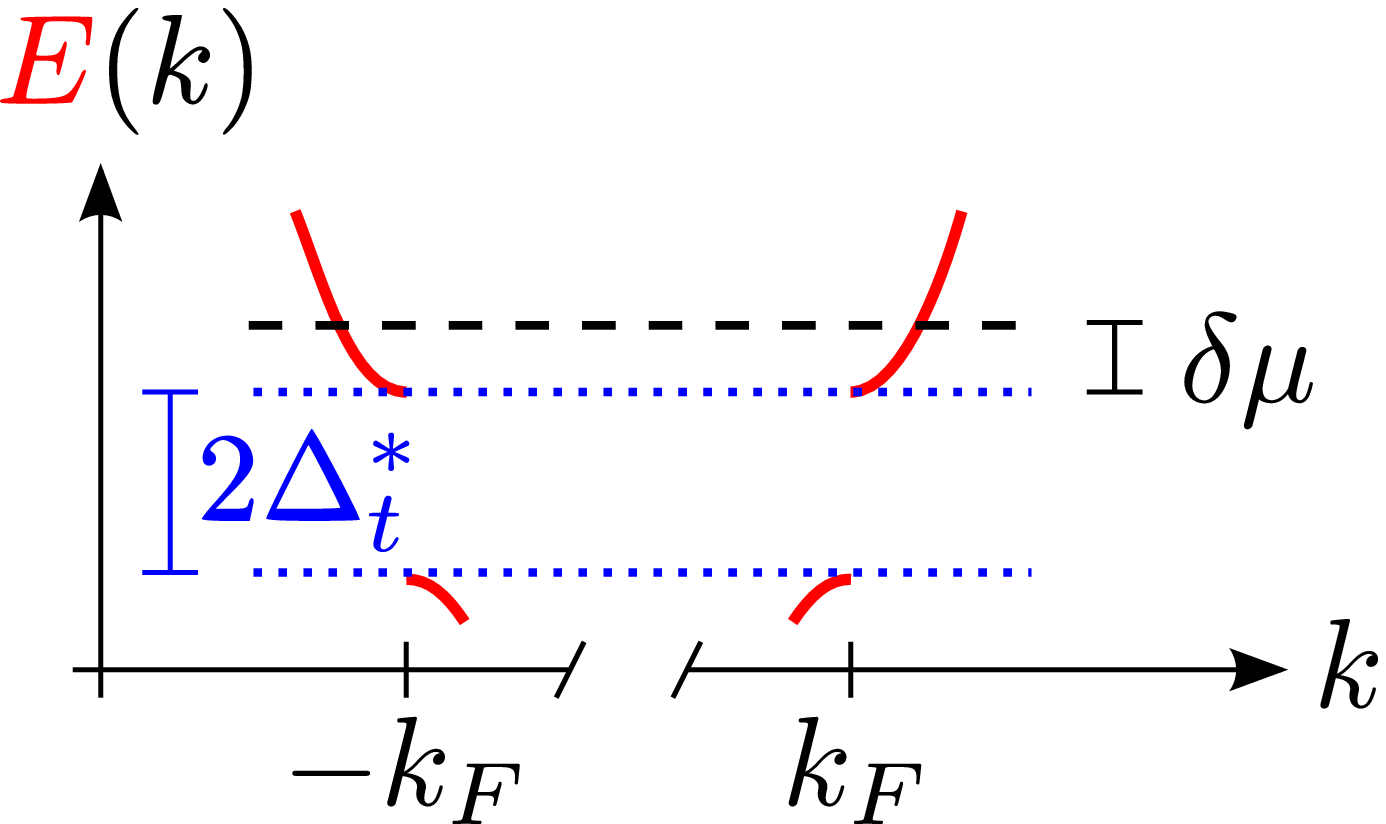}\\[0.1cm]~\\\raisebox{4.7cm}{(b)}\quad\includegraphics[scale=0.8]{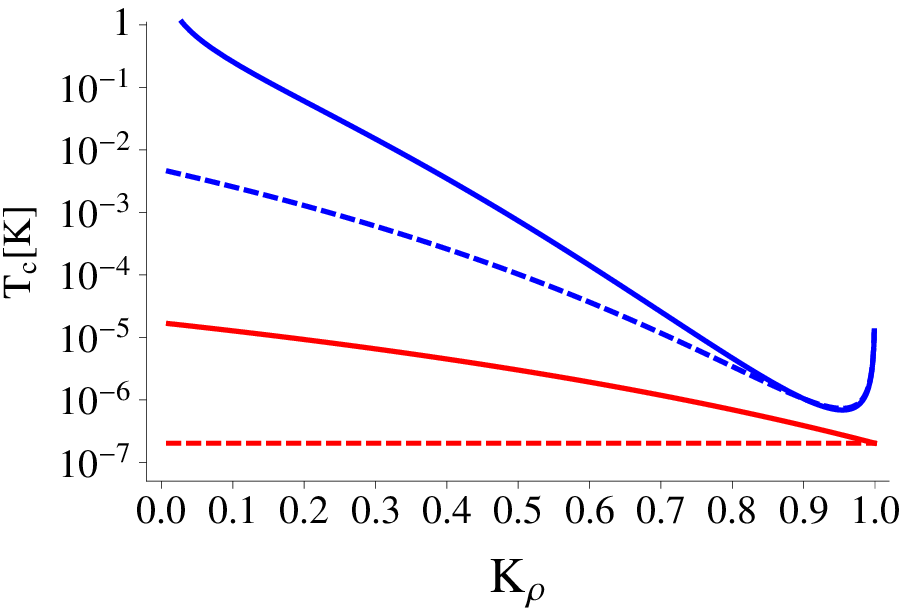}
\caption{(Color online) Panel (a): Energy $E$ of a stripe as a function of momentum $k$ (solid lines). The chemical potential is placed at a value of $\delta\mu$ above edge of the quantum Hall gap, whose width is $2\Delta_t$. The dotted lines mark the gap edges, the dashed line depicts the chemical potential. Panel (b): critical temperature of the nuclear spins for a stripe in GaAs with $\delta \mu=0.01$ meV (even an order of magnitude change of this parameter has a hardly visible effect in the current figure).  We also used $K_s=1$, $A\approx 90\,\mu\rm{eV}$ [\onlinecite{paget}], $I=3/2$, $a=5.63\,\mathring{\text{A}}$, $v_F=2\cdot10^5 \,\text{m/s}$, and wire cross section 80 nm$^2$, corresponding to $N_\perp \approx 2000$. The blue and red lines correspond to setting $\epsilon(m)=\epsilon_M$ and $\epsilon(m)=\epsilon_P+\epsilon_K$ in Eq.~\eqref{eq:MPK}, respectively. The solid (dashed) lines correspond to results with (without) the electron-nuclear feedback. The topmost line is thus $T_c$ calculated according to Eq.~\eqref{eq:Tcresult}.}
\label{fig:Tc}
\end{figure}

On a quantum Hall plateau, the system exhibits a full bulk gap for the electrons. As we argue in Sec.~\ref{sec:IIIB}, this gap strongly suppresses the RKKY exchange mediated by the electrons, which leads to unobservably small critical temperatures for nuclear spin order. If this order is to be established, the quantum Hall gap must drop to allow for a stronger RKKY exchange, and in turn for a higher $T_c$. This is indeed the case at the quantum Hall plateau edge, where the gap closes eventually because the interstripe tunneling, described by $\mathcal{H}_y$, becomes non-resonant. We analyze the nuclear spin order at the edge of the prototypical spin unpolarized quantum Hall plateau associated with the filling factor $\nu=2$. Close to, but outside the gap, the tunneling can still modify (bend) the dispersion despite its off resonant character. We therefore take the tunneling partially into account by the use of a modified Fermi velocity, and subsequently treat electron-electron interactions, and the hyperfine 
coupling within a given stripe, in a bosonized language. This yields a Luttinger liquid Hamiltonian density 
\be
\mathcal{H}_x=\frac{\hbar}{2\pi}\sum_{n,\kappa=\rho,s} \frac{u_\kappa'}{K_\kappa} (\partial_x \phi_{n,\kappa})^2+u_\kappa' K_\kappa (\partial_x \theta_{n,\kappa})^2,
\ee
where the charge and spin excitations have velocities $u_\kappa' = v_F'(\Delta_t)/K_\kappa$ with $v_F'(\Delta_t)=v_F\sqrt{1-\Delta_t^2/(\Delta_t+\delta\mu)^2}$, and where $\delta\mu$ denotes the chemical potential measured from the gap edge, see Fig.~\ref{fig:Tc}(a), and where $\Delta_{t}$ is the QHE gap opened by the tunneling $t_y$. If the nuclear order is not established (that is, at temperatures higher than $T_c$ calculated below), the RKKY exchange is given by Eq.~\eqref{eq:J gapless} upon replacing $v_F\to v_F^\prime(\Delta_t)$. Once the nuclear order is established (relevant for the calculation of the critical temperature), the feedback effects renormalize the velocity, $v_F^\prime(\Delta_t) \to v_F^{\prime \prime} =v_F^\prime(\Delta_t) \sqrt{(1+K_\rho^2)/(K_\rho^2+K_\rho^2K_s^2)}$, and the exponent, $g\to g^{\prime \prime} =2\,K_\rho/\sqrt{(1+K_\rho^2)(1+K_s^2)}$, see Appendix \ref{app:A2} for he derivation of these formulas. The RKKY exchange furthermore acquires an additional factor of $1/2$ because only the gapless electrons contribute resonantly to exchange [\onlinecite{braunecker_nuclear_order_prb_09}]. With these adjustments, we use Eq.~\eqref{eq:J gapless} in Eq.~\eqref{eq:Tc final} and obtain the critical temperature
\be
k_B T_c = \left(  \frac{1}{3N_\perp} A^2 I^{2} \left[ \frac{\hbar v_F^{\prime\prime}}{a} \right]^{1-2g^{\prime\prime}} c(g^{\prime\prime})   \right)^{\frac{1}{3-2g^{\prime\prime}}},
\label{eq:Tcresult}
\ee
which we plot in Fig.~\ref{fig:Tc}. As a main difference to Ref.~[\onlinecite{braunecker_nuclear_order_prb_09}], where only magnons along the stripe axis were considered, our calculation including also nuclear spin excitations within the cross sections thus results in an additional reduction of $T_c$ by a factor of $N_\perp^{-1/(3-2g '')}$. Still, we find that depending on the interaction strength controlling the power of $N_\perp$, the critical temperature can reach hundreds of mK. 

 \subsection{Experimental signatures}\label{sec:IIIE}
The direct experimental observation (in the form of a spatial image) of the periodic electronic stripes and/or nuclear helices [\onlinecite{atoms_exp,Yazdani_Shiba}] is not straightforward as the two dimensional electron gas is buried below the material surface. The same holds true for resistively detected NMR techniques [\onlinecite{muraki,smet_Nature,smet_PRL}], since the helical field averages out to zero along any direction. We therefore now propose several indirect indications which could establish the existence of the predicted effects.

{\it Modification of the Zeeman splitting.} When the nuclear spins are not ordered, they show an average uniform thermal magnetization of about 10\%-20\% parallel to the applied field [\onlinecite{chesi_thermal}]. The Overhauser field produced by this thermal polarization acts on the electrons as an additional Zeeman field. The thermal polarization is, however, destroyed when the nuclear spin helix forms. The Overhauser field produced by the helix is oscillatory and therefore averages out along any fixed direction. Depending on the material dependent relative signs of the electronic and nuclear $g$ factors and the hyperfine interaction, we expect that a helical nuclear spin polarization results in an effective reduction or enhancement of the Zeeman field seen by the electrons as compared to the paramagnetically ordered nuclear spin state.

\begin{figure}
\centering
\includegraphics[scale=0.3]{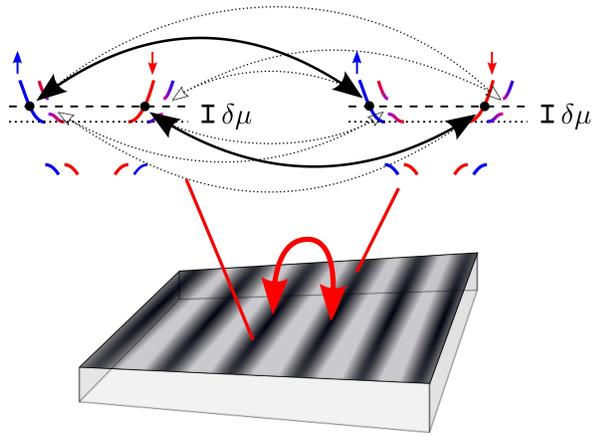}
\caption{(Color online) Illustration of the resistance increase below the ordering temperature $T_c$ (the degenerate spin up and down are offset for visibility). The opening of chiral gaps in the electron spectrum due to the formation of nuclear spin order in neighboring stripes (darker areas) suppresses the transport, both along each stripe (since half of the spectrum is gapped), and perpendicular to the stripes (because per spin, three out of four possible low energy tunneling processes are suppressed). The remaining interstripe tunnelings are shown as solid lines, the suppressed ones are depicted with dashed lines.}
\label{fig:conduct}
\end{figure}

{\it Increase of the resistance.} 
At the edges of a quantum Hall plateau, momentum conserving tunneling between neighboring stripes is weaker the more off-resonant the associated momentum kick becomes. As a consequence, momentum non-conserving tunneling events, allowed for instance due to the presence of impurities, can become important [\onlinecite{mcdonald_00}]. Let us first consider stripes of fixed orientation. Parallel to the stripe direction, transport is insensitive to the degree of detuning from the resonance as long as the system remains in the gapless phase, and as long as the stripes remain stable. When the nuclear spins order, half of the electron spectrum becomes gapped. In analogy to the resistance increase associated with a helical nuclear spin order in isolated quantum wires, the resistance $\rho_\parallel$ parallel to the stripes is then increased by roughly a factor of two [\onlinecite{braunecker_nuclear_order_prl_09,braunecker_nuclear_order_prb_09,meng_nuclear_order_double_wire}]. The resistance $\rho_\perp$ in the direction perpendicular to the stripes should be enhanced even stronger (by how much depends on the tunneling matrix elements associated with the hopping processes shown in Fig.~\ref{fig:conduct}). There, three out of four possible low energy tunneling processes per spin between neighboring stripes are suppressed by the opening of the partial gap in the electron system, see Fig.~\ref{fig:conduct}. In an experiment, the anisotropic resistance increase can, however, be masked in systems where the stripe orientation it not fixed. If the stripes are for instance aligned along (perpendicular) the current flow, a resistance measurement will will always yields $\rho_{\parallel}$ ($\rho_\perp$).

{\it NMR probing.}  The helical nuclear order can be probed in the NMR setup with two perpendicular magnetic fields, $B$ which is static and $B_{\rm osc}$ which oscillates at frequency $\omega$. Under resonance condition $\mu_N B = \hbar \omega$, standard for paramagnets and ferromagnets, the field will excite nuclear spins which are loosely bound to others (e.g. between stripes, on stripes surfaces, in stripes without order, etc). By nuclear diffusion, this tends to destroy any order, including the helical one. The resulting reduction of the nuclear polarization (from $m$ to $(1-p)m$) will reflect itself in a decreased transition temperature. If we assume that the equilibration times of the nuclear spins in the helical state and the collinear paramagnetic state are comparable, the NMR will have comparable effects on the reduction of the helical order. If we roughly describe such a reduction by reducing the effective density of nuclear spins taking part in the ordered state, $\rho_0\to (1-p)\rho_0$, which corresponds to reducing $A$ by the same amount, Eq.~\eqref{eq:Tcresult} gives the critical temperature reduced by a factor of $(1-p)^{2/3-2g''}$. Concluding, the expected reduction of $T_c$ is of the same order as $p$.

Interestingly, the nuclear spins inside stripes with helical order, firmly bound together by the RKKY interaction, cause additional response, at the resonance frequency given by the internal field $\hbar \omega = \mu_N B_{\rm eff} \equiv \epsilon_M$. Namely, because of the extremely narrow shape of the RKKY exchange, magnons with wavelengths of order ten microns are already "short wavelength" in our nomenclature (meaning dispersion-less, with energy $\epsilon_M$). A slight momentum offset of the oscillating NMR field due to even a very weak spin-orbit interaction in the electronic system then causes the helix to absorb at an NMR frequency set by the internal field $B_{\rm eff}$, rather than the external field $B$. The power-law temperature dependence of this internal field would be a clear sign of a nuclear helix. See Ref.~[\onlinecite{stano2014:U}] for details.

\subsection{Nuclear order in a gapped stripe}
\label{sec:IIIB}
Let us finally comment on the RKKY exchange on the quantum Hall plateau, where the system is fully gapped. The chosen gauge for the magnetic field allows us to exploit the translational invariance along $y$ by Fourier transforming the coordinate $n$ into the momentum $k_y$. In the non-interacting Eq.~\eqref{eq:Hx}, this leads to a simple index change $n \to k_y$, whereas Eq.~\eqref{1d_y} takes the form
\begin{equation}
\mathcal{H}_{y} = \sum_{k_y,\sigma} t_{y} e^{i(\phi(x)- k_y a_y)}\Psi^\dagger_{ k_y,\sigma}(x)  \Psi_{ k_y,\sigma}(x) + {\rm H.c.}~,\label{eq:tunneling_ham_fermi}
\end{equation} 
block diagonal in $k_y$ index. The bosonized form is
\begin{align}
\mathcal{H}_y&= \sum_{k_y,\sigma} \frac{t_{y}}{2\pi a}\label{eq:tunneling_ham_bos}
\Bigl[e^{i(\phi(x)-k_ya_y-2k_Fx)} e^{i\sqrt{2}(\phi_{k_y,\rho}+\sigma\phi_{k_y,s})}\nonumber\\
&\times e^{i(\phi(x)-k_ya_y+2k_Fx)}e^{-i\sqrt{2}(\phi_{k_y,\rho}+\sigma\phi_{k_y,s})}\Bigl]+{\rm H.c.},
\end{align} 
where Klein factors have been dropped. One can check that the backscattering terms associated with the interstripe tunneling $t_y$ and the intrastripe Overhauser field $B_{Ov} = 2mASI$ do not commute, expressing the fact that they correspond to competing intrastripe and interstripe orders, see Sec.~\ref{sec:IIID} below. On the plateau, where the tunneling is resonant, we assume $t_y\gg B_{Ov}$. The fate of the system can then be captured by a renormalization group (RG) analysis, in which the tunneling $t_y$ constitutes a relevant perturbation to $\mathcal{H}_x+\mathcal{H}_y$, and drives the system to the fully gapped strong coupling fixed point associated with the interstripe tunneling. The presence of a gap does not, however, exclude the possibility to form a nuclear spin ordered state: even a fully gapped electron system can mediate an RKKY interaction of reduced strength [\onlinecite{klinovaja_sc,braunecker_sc}].

The RKKY exchange can in principle be calculated upon integrating the RG flow of the interstripe tunneling until the tunneling gap $\Delta_t$ associated with $t_y$ reaches the bandwidth, and subsequently expanding the sine-Gordon term in Eq.~\eqref{eq:tunneling_ham_bos} to second order. Because, however, the  intrastripe Coulomb repulsion $\mathcal{H}_C$ given in Eq.~\eqref{eq:coulomb_ferm}, as well as Eq.~\eqref{eq:llham1}, are highly off-diagonal in $k_y$ space, we were not able to find an analytical solution for the RKKY exchange on the quantum Hall plateau.

\begin{figure}
\centering
\includegraphics[scale=0.9]{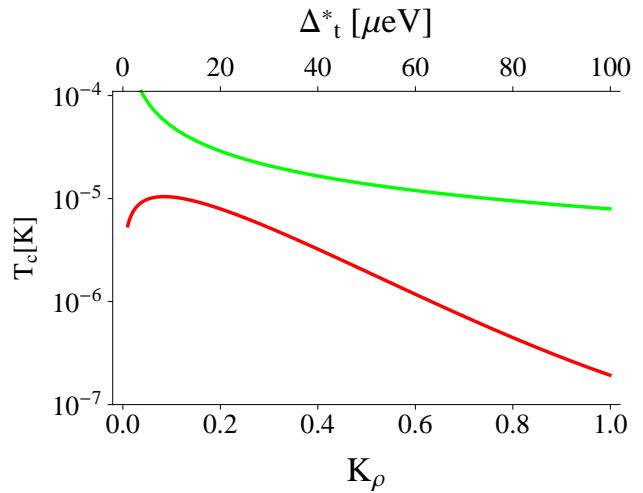}
\caption{(Color online) Critical temperature of the nuclear order in a single gapped stripe with intrastripe backscattering as a function of interaction strength $K_\rho$ (red (bottom) line, bottom x axis) at fixed value of the gap $\Delta_{t}^*=0.1$ meV, and as a function of the gap (green (upper) line, upper x axis) at a fixed interaction parameter $K_\rho=0.3$. All other parameters are the same as those used in Fig.~\ref{fig:Tc}.  As discussed in Sec.~\ref{sec:IIIB} and Appendix \ref{append:rkky_gapped}, we expect the critical temperature of nuclear spin ordering on a quantum Hall plateau to have a similar order of magnitude.}
\label{fig:Tc gapped}
\end{figure}

In order to nevertheless get an impression of the order of magnitude of critical temperatures in a gapped system, we analyze the more simple case of intrastripe backscattering instead of the interstripe backscattering. While this process gives rise to a different state than the quantum Hall state, the suppression of the RKKY exchange due to the presence of a full bulk gap is expected to be qualitatively independent of the nature of this bulk gap. To set apart the results obtained within this model by notation, we use $\Delta_t^*$ for the bulk gap. Details of the calculation can be found in Appendix \ref{append:rkky_gapped}, where we estimate the critical temperature of nuclear spin order in a stripe gapped by intrastripe backscattering as

\be
J_{2k_F} \approx -\frac{A^2a}{4\pi v_F N_\perp^2}
\frac{1}{1-K_\rho}\left(\left(\frac{\hbar v_F\sqrt{K_\rho}}{\Delta_{t}^*a}\right)^{1-K_\rho}-1\right).
\label{eq:J gapped}
\ee
This equation is valid for temperatures much smaller than $\Delta_{t}^*$, the gap associated with the intra stripe backscattering. Assuming that we are in the regime where Eq.~\eqref{eq:Tc final} is valid (so that the magnon energy is dominated by the RKKY energy), we plot the critical temperature as a function of the interaction strength $K_\rho$ in Fig.~\ref{fig:Tc gapped} for typical GaAs parameters. As seen from there, the critical temperature is well below 1 mK, and therefore unobservably small. This illustrates the strong reduction of the RKKY exchange, and of the nuclear spin ordering temperature by the gap. By analogy, we thus expect the ordering temperature of nuclear spins of a strip of stripe in the gapped quantum Hall state to be outside the reach of current experiments.

\subsection{Disordered to ordered phase transition}\label{sec:IIID}
As we have already noted, the QHE gap and the partial gap from the electron-nuclear feedback are induced by coupling different electron branches. While the former arises from spin conserving momentum-inverting interstripe hopping, the latter results from spin-flipping momentum-inverting intrastripe backscattering. Because of this, the two gaps are incommensurate and the larger one strongly suppresses the smaller one. As a consequence, at temperatures below the nuclear order critical temperature, the transition between a fully gapped and a partially gapped stripe is abrupt. For $\Delta_t>\Delta$, where $\Delta$ is the partial gap in the electron system opened by the feedback of the helically ordered nuclear spins, the stripe is in the fully gapped phase (described in Sec.~\ref{sec:IIIB}) and thus without a nuclear order at experimentally relevant temperatures. Decreasing the QHE gap by moving the magnetic field strength off the resonance, the stripe suddenly jumps into the gapless phase at $\Delta_t \simeq \Delta$(Sec.~\ref{sec:IIIC}), where the nuclear order is established, thereby closing the QHE gap, and opening the partial gap in the electron system.

While a single stripe transition is abrupt, the bulk transition can still be gradual. Namely, when the magnetic field is moved away from resonance, the stripes can organize into a periodic superstructure, with unequal distances inside a supercell [\onlinecite{klinovaja_strip_of_stripes}]. This allows to keep the resonance condition for some stripes, enjoying the energy gain from open QHE gaps but not ordering the nuclear spins, whereas the remaining stripes are off-resonant for the tunneling, but establish a nuclear order. By changing the ratio between the number of stripes in these two sets, the transition is gradual. Importantly, this argument shows how the stripe phase, which is supported by the QHE gaps, can coexist with the nuclear order phase, which requires gapless stripes to achieve experimentally relevant critical temperatures. This also suggests that the nuclear polarization arises close to the plateau edge, rather than deep inside the plateau.

\section{Discussion of the physics of nuclear spin order}\label{sec:disc}
Let us now discuss the nuclear spin order from broader perspective. The order arises due to nuclear spin-spin interaction, mediated by free electrons, according to Eq.~\eqref{eq:RKKY}. On short distances, this interaction is ferromagnetic, irrespective of the sign of the electron spin-nuclear spin interaction (the Fermi contact interaction). Namely, a nuclear spin induces a spin polarization in the electron gas, which is seen by other nuclear spins as a source of energy through the same contact interaction. The helical form of the order, on the other hand, is a consequence of the long distance behavior of the electron spin polarization, which is oscillatory with the wave vector $2k_F$, in a complete analogy to the Friedel oscillations. This also means that such a helical order will arise only in one dimension.  In a helically ordered state, nuclear spins contribute constructively to the electron spin polarization, and a macroscopic helical Knight field is established. The critical temperature for the nuclear order is then given by the Zeeman energy of a nuclear spin in the Knight field. (The absence of lower energy excitations, which do not allow for an order in one dimension, is here due to, together, a finite size of the wire and a very singular shape of the spin susceptibility.) 

Due to the smallness of the Fermi contact interaction constant $A$, however, one expects a rather low critical temperature. This is indeed the case, since $T_c$ is of order $\mu$K for non-interacting electrons (demonstrated in Fig.~\ref{fig:Tc}b). Here is where the strong electron-electron interactions, typical for low-dimensional systems, are essential, pushing the critical temperature into experimentally observable values of up to a tenth of a Kelvin. In addition to the enhancement of the Knight field itself, the electron-electron interactions contribute to the energy gain of the ordered system. This way, the critical temperature can actually overcome the limit $T_{c0} \sim A (\rho_e/\rho_I)$ set by the maximal achievable Knight field, the latter for a fully spin-polarized electronic band (here $\rho_e$ and $\rho_I$ is the three dimensional density of electrons and nuclear spins, respectively).

The very strong (many magnitudes) enhancement of the critical temperature through electron-electron interactions signals a phase transition in the electronic system itself. Formally, it shows up as a divergence of the electron response in the zero temperature limit. Such a phase transition (for which the helically ordered nuclear spins serve as a symmetry breaking field) can arise only in the presence of electron-electron interactions, where a spin or charge density wave can support itself in the system ground state, e.g., by the Peierls mechanism.

Finally, we note that this work ignores how the thermodynamic equilibrium is achieved. However, it is an experimentally well established fact that the electrons are dominant in providing the dissipation channel for nuclear spins. Namely, the nuclear spin order can be maintained over hours or days if the hyperfine interaction is absent, even at room temperature [\onlinecite{mcneil1976:PRB,ono2004:PRL,ren2010:PRB}]. Therefore, spin polarization of electrons will have strong influence on dynamics of the nuclear order. Assume a partial helical order in nuclear spins is established, triggering a strong spin polarization in the electronic system due to interactions. This electron spin polarization will serve as a source for the nuclear spin order, in an analogy to dynamical nuclear spin polarization, routinely observed in the quantum Hall regime [\onlinecite{dixon1997:PRB,kou2010:PRL,wald1994:PRL}]. Second, with a full electron spin order there is no channel for the nuclear spins to decay into a disordered state, even if, e.g., the temperature is taken above the critical temperature. Because of these possible dynamical effects, we expect that the nuclear order can actually be established at temperatures even higher than those we have calculated here, especially in experiments involving electronic transport. We will investigate these dynamical mechanisms of nuclear order in a future work.

\acknowledgements
We acknowledge stimulating discussions with B. Frie{\ss}, and thank K. von Klitzing for drawing our attention to the problem of nuclear spin order in quantum Hall systems. We acknowledge support by the SNF, NCCR QSIT, and Harvard Quantum Optical Center. P.~S. acknowledges the support of QIMABOS-APVV-0808-12.

\appendix

\section{Zeeman Hamiltonian}\label{sec:append_ham}
The Zeeman effect lifts the degeneracy between spin up and spin down, which in turn gives rise to distinct Fermi velocities for the two spin species. In order to analyze the impact of this velocity difference on our results, we linearize Eq.~\eqref{eq:Hx} around the Fermi points at momentum $\pm k_{F\uparrow,\downarrow}$ written in term of right and left mover fields, 
\begin{equation}
\mathcal{H}_{x}=\sum_{n, \sigma=\uparrow, \downarrow} \hbar \upsilon_{F\sigma} [R_{n,\sigma}^\dagger(-i\partial_z)R_{n,\sigma}-L_{n,\sigma}^\dagger(-i\partial_z)L_{n,\sigma}] .
\end{equation}
The Coulomb repulsion between electrons, which is added next, is strongest for electrons within the same stripe. Projecting the intrastripe interaction onto the Fermi points within each stripe, we obtain several matrix elements associated with a momentum transfer close to zero, $2k_{F\uparrow}$, $2k_{F\downarrow}$, and $k_{F\uparrow}\pm k_{F\downarrow}$. We retain only the dominant terms associated with zero momentum transfer, which are given in Eq.~\eqref{eq:coulomb_ferm}. Next, we introduce bosonic fields $\phi_{n,\sigma}$ and $\theta_{n,\sigma}$ which fulfill the standard commutation relations$[\phi_{n,\sigma} (x), \theta_{n,\sigma'}(x') ]=\delta_{\sigma,\sigma'}\delta_{n,n'}(i\pi/2) {\rm sgn}(x'-x)$, where $\phi_{n,\sigma}$ relates to the integrated density of particles of spin $\sigma$ in stripe $n$, while $\theta_{n,\sigma}$ is proportional to their integrated current density [\onlinecite{giamarchi_book}]. As a result, the right and left mover fields $r_{n,\sigma}$ with $r=R,L$ are represented as
\begin{align}
r_{n,\sigma}(x) = \frac{U_{rn\sigma}}{\sqrt{2\pi a}}\,e^{-i(r\phi_{n,\sigma}(x)-\theta_{n,\sigma}(x))}
\end{align}
where $U_{nr\sigma}$ is a Klein factor, and $a$ is a short-distance cut-off. From there, the Hamiltonian density can be brought to a diagonal form using a new bosonic field basis $(\phi_\vartheta,\theta_\vartheta)$ detailed below, which yields
\begin{align}
 \mathcal{H}_{x} &=\frac{\hbar}{2\pi}\sum_{n,\kappa=\pm}u_\kappa(\partial_x \phi_{n,\kappa})^2+u_\kappa (\partial_x \theta_{n,\kappa})^2.
\end{align}
The effective velocities are given by
\begin{align}
u_{\pm} &= \sqrt{\frac{u_{\uparrow}^2+u_{\downarrow}^2}{2}\pm\sqrt{ \left(\frac{u_{\uparrow}^2-u_{\downarrow}^2}{2}\right)^2 + U^2 v_{F\downarrow} v_{F\uparrow} }},
\end{align}
where we use the notations $u_{\uparrow,\downarrow}= v_{F\uparrow,\downarrow}\sqrt{1+(U/v_{F\uparrow,\downarrow})}$. The new basis is given by
 \begin{subequations}
 \begin{align}
\phi_{n,\uparrow}&= \sqrt{u_{\uparrow}K_{\uparrow}}\left( \frac{\cos\beta}{\sqrt{u_{+}}}\,\phi_{n,+}+\frac{\sin\beta}{\sqrt{u_{-}}}\phi_{n,-}\right),\\
\phi_{n,\downarrow}&= \sqrt{u_{\downarrow}K_{\downarrow}}\left( - \frac{\sin\beta}{\sqrt{u_{+}}}\,\phi_{n,+}+\frac{\cos\beta}{\sqrt{u_{-}}}\phi_{n,-}\right),\\
\theta_{n,\uparrow}&= \frac{1}{\sqrt{u_{\uparrow}K_{\uparrow}}}\left[ \sqrt{u_{+}}(\cos\beta) \phi_{n,+}+\sqrt{u_{-}}(\sin\beta) \phi_{n,-}\right],\\
\theta_{n,\downarrow}&= \frac{1}{\sqrt{u_{\downarrow}K_{\downarrow}}}\left[ - \sqrt{u_{+}}(\sin\beta) \phi_{n,+}+\sqrt{u_{-}}(\cos\beta) \phi_{n,-}\right],
 \end{align}
 \label{eq:diagonalization_trafo}
 \end{subequations}
with $\tan(2\beta) = 2 U \sqrt{v_{F\uparrow}v_{F\downarrow}}/(U+v_{F\uparrow}+v_{F\downarrow})(v_{F\downarrow}-v_{F\uparrow})$. This basis differs from the usual spin and charge basis, with $\phi_{n,\rho} = (\phi_{n,\uparrow}+\phi_{n,\downarrow})/\sqrt{2}$, $\phi_{n,s} = (\phi_{n,\uparrow}-\phi_{n,\downarrow})/\sqrt{2}$, and the conjugate fields $\theta_{n,\rho/s}$ because the Zeeman effect leads to a coupling between the spin and charge fields proportional to $v_{F\uparrow} - v_{F\downarrow}$. When the Zeeman effect is neglected by setting $v_{F\uparrow}= v_{F\downarrow}= v_{F}$, the velocities $u_{\pm}$ come back to standard velocities in the charge and spin sectors, $u_+=u_\rho=v_{F}\sqrt{1+(2U/v_{F})}$ and $u_-=u_s=v_{F}$. Similarly, the fields then obey $\phi_{n,+} = \phi_{n,\rho}/\sqrt{K_\rho}$, and $\phi_{n,-}=\phi_s/\sqrt{K_s}$ with $K_i=v_F/u_i$, as well as  $\theta_{n,+} = \theta_{n,\rho}\,\sqrt{K_\rho}$, and $\theta_{n,-}=\theta_{n,s}\,\sqrt{K_s}$. 

Let us now analyze the importance of the Zeeman effect for our results. In the experimentally relevant magnetic field range of up to a few Tesla, we have checked that the velocity difference, and the resulting coupling between spin and charge are in fact not important for the nuclear spin order discussed in this work. To illustrate this finding, we recall that the spin resolved Fermi velocities are given by $v_{F\sigma} = v_F \,\sqrt{1 -\sigma \Delta_Z/\mu}\approx v_F(1-\sigma\Delta_Z/2\mu)$. Together with the fact that the coupling between spin and charge is proportional to $v_{F\uparrow}- v_{F\downarrow}$, this implies that the RKKY exchange calculated in this work exhibits only negligibly small corrections $\sim (\Delta_Z/\mu)^2$ due to the presence of a Zeeman splitting.

\section{Critical temperature}
\label{app:Tc}
Here we calculate the magnon spectrum and the resulting critical temperature in the $n$-th stripe, the index of which is omitted in this section.

\subsection{Magnon energies}
\label{app:Tc1}
Here, we provide details for obtaining the critical temperature discussed in Sec.~\ref{sec:Tc}. We consider quasi one-dimensional electrons inside a given stripe, with wave-functions factorized into longitudinal and transverse components, along $x$, and $(y,z)$ directions, respectively. Dropping the stripe index $n$, we approximate the lowest transverse subband wave-function $\phi_0(y,z)$ by a constant for coordinates $(y,z)$ inside the stripe, $\phi_0(y,z)=1/\sqrt{C}$, and zero otherwise. Although with minor consequences on physics, this is a huge technical simplification, which renders the problem one dimensional. We split the wire along its axis into cylinders of length $a$ centered at $x_i$, referred to as transverse planes and labeled by the index $i=1,\dots, N$, with $N=L/a$. If the nuclear spin volume density is $\rho_0$, there are $N_\perp=Ca\rho_0$ nuclear spins within a volume corresponding to a transverse plane. Its total spin operator is
\begin{equation}
{\bf \tilde{I}}_{i} = \sum_{l\in i} {\bf I}_l,
\end{equation}
with which we write the RKKY Hamiltonian, Eq.~\eqref{eq:RKKY}, as
\begin{equation}
H_R=\sum_{i, j} {\bf \tilde{I}}_{i} \cdot {\bf \tilde{I}}_{j} J_{i j}.
\label{eq:RKKY2}
\end{equation}
We choose the basis of a transverse plane to be the set of states labeled by the total spin $L_i$, its projection along the local order direction axis, $M_i$, and an additional quantum number $\xi_i=1,\ldots,\Xi(L_i)$. A fully ordered transverse plane has $L_i=M_i=N_\perp I$, and there is just a single such state, $\Xi=1$, with all constituent spins collinear. For smaller $L_i$ there are more states differing in their symmetry with respect to pairwise swaps of constituent spins. The basis of the total system is a tensor product of bases of individual transverse planes.

The Hamiltonian in Eq.~\eqref{eq:RKKY2} preserves both $L_i$ and $\xi_i$ quantum numbers of each plane. We therefore fix the set of numbers $\{ L_i, \xi_i \}_{i=1}^N$, and diagonalize the RKKY Hamiltonian within this subspace using the Holstein-Primakoff ansatz. To this end, we first introduce a linear transformation of the spin operators which undoes the rotation in Eq.~\eqref{eq:ground state},
\be
\hat{\bf I}_i = R_{{\bf h}, 2k_F x_i}^{-1} \tilde{\bf I}_i.
\ee
Inserting this into the RKKY Hamiltonian gives
\be
H_R  = \sum_{ij} \hat{\bf I}_i \cdot \hat{\bf J}_{ij} \cdot  \hat{\bf I}_j,
\label{eq:helical RKKY}
\ee
where the transformed RKKY exchange is
\be
\hat{\bf J}_{ij} 
=  \left( 
\begin{tabular}{ccc}
$\cos [2k_F(x_i-x_j)]$ & $\sin [2k_F(x_i-x_j)]$&0\\
$-\sin [2k_F(x_i-x_j)]$ & $\cos [2k_F(x_i-x_j)]$&0\\
0 & 0 & 1
\end{tabular} 
\right) J_{ij}.
\label{eq:helical RKKY2}
\ee
In this coordinate system, the ground state (of the Hilbert subspace) corresponds to $\hat{\bf I}_i = (L_i,0,0)$ and is therefore suitable for the Holstein-Primakoff ansatz through the following substitutions
\begin{eqnarray}
\label{eq:HPreal1}
\hat{I}^x_j &=& L_j - n_j.\\
\hat{I}^y_j &=& \frac{1}{\sqrt{2}} \left( a_j^\dagger \sqrt{L_j-n_j/2} + \sqrt{L_j-n_j/2}\,a_j \right),\\
\hat{I}^z_j &=& \frac{i}{\sqrt{2}}  \left( a_j^\dagger \sqrt{L_j-n_j/2} - \sqrt{L_j-n_j/2}\,a_j \right).
\label{eq:HPreal3}
\end{eqnarray}
The bosonic operators for transverse planes, $a_i$, fulfill standard commutation relations $[a_i,a^\dagger_j]=\delta_{ij}$, zero otherwise, and $n_j=a^\dagger_j a_j$.

The standard procedure is a calculation in Fourier space that treats the higher order bosonic terms in some approximation scheme. Namely, the representation in Eqs.~\eqref{eq:HPreal1}-\eqref{eq:HPreal3} is exact, but complicated to use because of the square roots. To proceed, we parametrize the value of $L_j$ by $L_j=N_\perp I-D_j$ and Taylor expand the square roots in the ratio $(D_j+n_j/2)/N_\perp I \leq 1$. In terms containing more than two bosonic operators, we employ the mean field approximation by replacing a creation-annihilation operator pair by its expectation value $a_i^\dagger a_j \to \langle n_i \rangle \delta_{ij}$. In terms of these variables, the magnetization $m$, defined in Eq.~\eqref{eq:ground state}, is given by $N N_\perp I (1-m) = \sum_j D_j + \langle n_j \rangle$.  Finally, we introduce discrete Fourier transforms between the real space and momentum variables according to   
\be
f_q = \sum_j \exp(-i q x_j) f_j, \quad f_j = \frac{1}{N} \sum_q \exp(i q x_j) f_q.
\label{eq:FTdefinition}
\ee
A short calculation along these lines gives
\begin{align}
\hat{I}_q^x &= \delta_{q,0} \left(N N_\perp I -D_0-\frac{1}{N}\sum_p a_p^\dagger a_{p} \right)\nonumber\\
& -(1-\delta_{q,0}) \left(D_q + \frac{1}{N}\sum_p a_p^\dagger a_{p+q}\right)
\label{eq:HPmomentum1}\\
\hat{I}_q^y &= \sqrt{\frac{m N_\perp I }{2}} \left( a^\dagger_{-q} + a_q \right)\nonumber\\
&- \frac{1}{2N \sqrt{2 N_\perp I }}\sum_{p\neq q} (a^\dagger_{-p}+a_p) D_{p-q},\\
\hat{I}_q^z &= i \sqrt{\frac{m N_\perp I }{2}} \left( a^\dagger_{-q} - a_q \right)\nonumber\\
&- \frac{i}{2N \sqrt{2 N_\perp I }}\sum_{p\neq q} (a^\dagger_{-p}-a_p) D_{p-q}.
\label{eq:HPmomentum3}
\end{align}
To arrive at these, we have replaced the lowest order Taylor expansion of $\sqrt{m}$ in the small parameter $1-m$ by the expression $\sqrt{m}$ itself. We have also split the results into those not containing, and containing $D_{q\neq 0}$ terms, respectively. Compared to the former, the latter give a negligible contribution in the final result. This is so because, first, since the numbers $D_i$ are all positive, it holds that $D_0\geq |D_{q\neq 0}|$. For a typical configuration at an incomplete magnetization, $m<1$,  $D_0 \gg |D_{q\neq 0}|$ holds the better the larger $N$ is. Second, these terms represent electron scattering on the spatial structure of transverse planes total spins $L_i$, and on thermally excited magnons [for the last term of Eq.~\eqref{eq:HPmomentum1}]. Such terms break the translational symmetry of the problem (within the given Hilbert space subspace) and greatly complicate the analysis. However, in calculating the statistical sum over all possible configurations, which is our final goal, 
we expect such contributions to average out, so that we neglect them already at this point.

The Fourier transform of the RKKY tensor comes straightforwardly from Eqs.~\eqref{eq:helical RKKY2} and \eqref{eq:FTdefinition} as
\be
\hat{\bf J}_q =
\left(
\begin{tabular}{ccc}
$J_q^+$ & $-i J_q^-$  & 0\\
$i J_q^-$ & $J_q^+$ & 0\\
0 & 0 & $J_q$
\end{tabular} 
\right),
\label{eq:RKKYmomentum}
\ee
with $J_q$ the Fourier transform of the original RKKY exchange $J_{ij}$ and $J_q^\pm = (J_{q+2k_F} \pm J_{q-2k_F})/2$.

We now insert Eqs.~\eqref{eq:HPmomentum1}-\eqref{eq:HPmomentum3}, neglecting the second terms on the right hand sides (within this approximation, the off-diagonal terms of the tensor in Eq.~\eqref{eq:RKKYmomentum} do not contribute to finite momenta magnons), and Eq.~\eqref{eq:RKKYmomentum} into Eq.~\eqref{eq:helical RKKY}.  With the mean field approximation for higher order terms, we obtain a Hamiltonian bilinear in bosonic operators, which we diagonalize by a Bogoliubov transformation, finally arriving at
\be
H |_{{\{L_i,\xi_i\}}_{i=1}^N} = \frac{1}{N} \Big(\sum_i L_i \Big)^2 J_{2k_F} + \sum_q \epsilon_q b_q^\dagger b_q.
\label{eq:RKKY final}
\ee
The bosonic operator $b_q^\dagger$ creates a magnon with momentum $q$ and energy
\be
\epsilon_q = 2 m N_\perp I \sqrt{(J_q^+ - J_{2k_F})(J_q - J_{2k_F})}.
\label{eq:magnon dispersion}
\ee 
The magnon spectrum is gapless at three points, $q_0=0,\pm 2k_F$, each of which corresponds to a Goldstone mode of a global spin rotational symmetry. We find three Goldstone modes, as our model has three axes of rotational symmetry: two for a rotation of the helical plane vector ${\bf h}$, and one for a rotation of the helix within the plane, ${\bf I_0}$ around ${\bf h}$. Energy at momentum $q$ around each Goldstone mode momentum $q_0$ (we denote $\delta q = |q-q_0|$) is obtained in the lowest order by a Taylor expansion as 
\be
\epsilon_{\delta q \ll q_w} \approx  m N_\perp I \sqrt{ -2 J_{2k_F} (\partial_{kk} J_{2k_F} ) } \,\, \delta q \equiv \hbar c \delta q,
\label{eq:long wavelength}
\ee 
(three identical copies of) a linear spectrum of magnons with velocity $c$. The validity of this expansion defines the long wavelength magnons, through the momentum $q_w$. For all other momenta, where $|J_q| , |J^+_q| \ll |J_{2k_F}|$, we get the spectrum of short wavelength magnons as
\be
\epsilon_{q} \approx - 2 m N_\perp I J_{2k_F} \equiv \epsilon_R.
\label{eq:epsilon R}
\ee 
For completeness, we note that this expansion is not valid at $q=\pm 4k_F$, where $\epsilon_{\pm 4k_F} = \epsilon_R/\surd{2}$. However, since here the spectrum is gapped by an energy comparable to $\epsilon_R$, the mistake we do by replacing the magnon energies by $\epsilon_R$ in this small region of momenta is completely irrelevant for the statistical sum evaluation.

\begin{figure}
\includegraphics[width=0.4\textwidth]{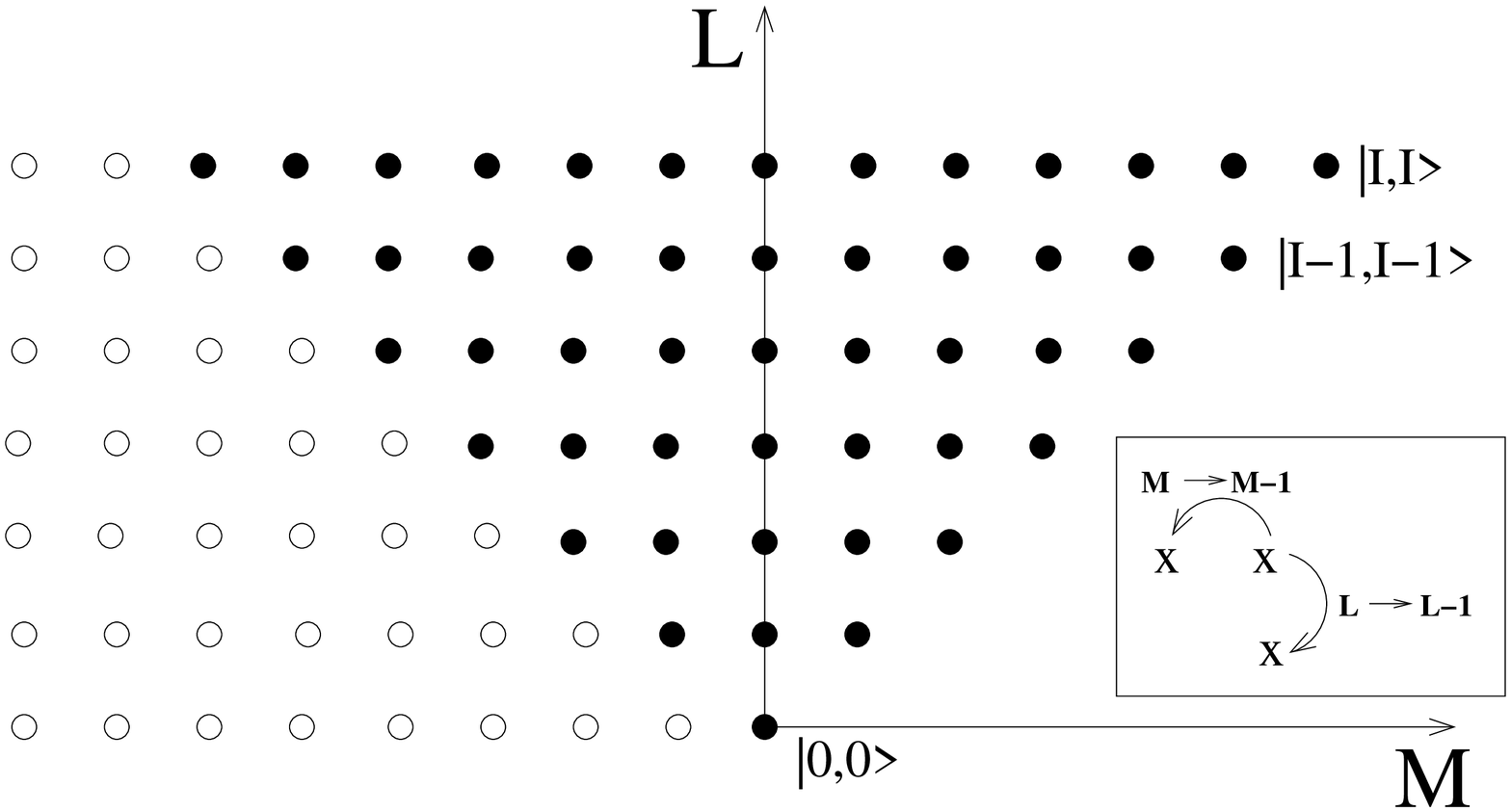}
\caption{Hilbert space of a transverse plane after a mean field approximation of the Holstein-Primakoff ansatz. Each state is labeled by the total spin $L$ (vertical axis), the spin projection along certain axis, $M$ (horizontal axis) and the exchange symmetry structure (label $\xi$; not given). The filled dots represent physical states. The empty dots are fake states introduced by the approximate (unrestricted) Holstein-Primakoff representation. In such a representation, each line starts with a ``seed'' state (the rightmost), with the boson ladder extending to the left. Exciting the boson to a higher state shifts the position to the left by one step, keeping $L$ and $\xi$ the same.}
\label{fig:Hilbert space} 
\end{figure}

Let us now come back to the first term in Eq.~\eqref{eq:RKKY final}. It describes the energy dependence on the subspace quantum numbers $\{L_i, \xi_i\}$, and can be written as
\begin{equation}
E(\{L_i\}) \equiv \frac{1}{N} \Big(\sum_i L_i \Big)^2 J_{2k_F} = -\frac{1}{2} \Big(\sum_i L_i \Big)  \epsilon_R.
\end{equation}
This shows that a decrease of the total spin of any given transverse plane, $L_i\to L_i-1$, costs the same energy as the excitation of a short wave-length magnon, $\epsilon_R$, given in Eq.~\eqref{eq:epsilon R}. These two types of excitations are illustrated in Fig.~\ref{fig:Hilbert space}.

The partition function can be now written as
\begin{equation}
\mathcal{Z} = \bigotimes_{i=1}^N\sum_{L_i=N_\perp I}^{-N_\perp I} \sum_{\xi_i=1}^{\Xi(L_i)} e^{ -\beta E(\{L_i\})}  \prod_q {\sum_{n_q}}^\prime e^{ -\beta n_q \epsilon_q}.
\label{eq:Z}
\end{equation}
We remind that the index $i=1,\ldots, N$ labels transverse planes, each with a total spin $L_i$ and a symmetry quantum number $\xi_i$. Within a subspace of fixed $\{L_i, \xi_i\}$, the states are specified by the set of excitation numbers $n_q$ of magnons with quantum numbers $q = 2\pi /L \times (0, \ldots, N-1) $, from which the Goldstone modes are excluded, $q\neq 0, \pm 2k_F$. The formula is, however, difficult to evaluate in its exact form. This is due to complicated restrictions on the magnon occupation number (see below), denoted by the prime of the summation through $n_q$, and the complicated degeneracy factors $\Xi(L_i)$. The latter can be roughly estimated by replacing a single nuclear spin $I$ by $2I$ spins 1/2, in which case
\begin{equation}
\Xi(L_i) = \binom{2N_\perp I}{D_i} - \binom{2N_\perp I}{D_i \pm 1 } \approx \frac{1}{D_i!}\left(N_\perp I\right)^D_i, 
\end{equation}
where the factor $\pm 1$ equals the sign of $-L_i$, the expansion holds for $D_i \equiv N_\perp I - L_i \ll N_\perp I$.

To proceed with Eq.~\eqref{eq:Z}, we first assume that the long wavelength magnons contribution to $\mathcal{Z}$ is negligible. This implies that their number is negligible compared to short wavelength magnons (the implication cannot be reversed). Replacing the energy of this negligible small set of excitations by $\epsilon_R$, we arrive at a remarkably simple result: it does not matter how the nuclear magnetization is diminished, the energy cost of any kind of spin flip is the same, and depends only on the magnetization $m$. The system differs from a set of independent spins in a magnetic field 
\be
\mu B (m) = -2 m N_\perp I J_{2k_F},
\ee
only by the basis. In our derivation of Eq.~\eqref{eq:RKKY final}, there was no limit on magnon occupations. We lost the proper restrictions, assured by Eqs.~\eqref{eq:HPreal1}-\eqref{eq:HPreal3}, by a mean field approximation of the Taylor expanded square root factors. Such an approximation results in no limit on the magnetization decrease within a subspace with given $\{L_i, \xi_i\}$, specifically, allowing for unphysical states with $\sum_q n_q  > \sum_i L_i$ (see the illustration in Fig.~\ref{fig:Hilbert space}). It is not simple to correct for this exactly, since the true restriction is $n_i \leq L_i$. However, by imposing the proper condition on average via the requirement $n_q \leq \sum_i L_i/N$, our system becomes {\it exactly equivalent} to non-interacting spins. Up to the negligibly small fraction of long wave-length magnons (compared to the total number of excitations), and postponing the proof that other energy costs depend only on the magnetization $m$ (which we do below), this finishes the way 
to Eq.~\eqref{eq:magnetization definition}.

Let us now consider the long-wavelength magnons. These are gapless, unlike the short wavelength magnons, so that upon lowering the temperature there is a smaller and smaller set of these with larger and larger occupations. To describe such a case, we restrict the phase space in Eq.~\eqref{eq:Z} to the only gapless subspace $L_i=N_\perp I$ for all $i$, for which $\Xi_i=1$ and only magnons with linear dispersion, $q\leq q_w$. Finally, ignoring the restrictions on the magnon population, $\mathcal{Z}$ becomes the partition function of a set on independent bosons, which gives Eq.~\eqref{eq:Tc2} as their total number.

We have thus derived two limits for the partition function in which the transition temperature can be calculated easily. These limits correspond to the long wavelength magnons being negligible, and dominant, respectively. In a general case, both short and long wavelength magnons contribute together, and the true transition temperature will be slightly lower than the minimum of the values calculated from Eqs.~\eqref{eq:magnetization definition} and \eqref{eq:Tc2}. 

We now illustrate the two limits by deriving the crossover wire length $L_{max}$, above which the long-wavelength magnons dominate. We define it as the wire length at which the long wavelength magnon population, given by Eq.~\eqref{eq:Tc2}, reaches the total magnetization $N N_\perp I$. Approximating the sum by an integral, and extending its upper bound to infinity, we get
\be
L_{\rm max} \approx -\frac{2\pi \hbar c}{k_B T} \ln^{-1} \left( 1- e^{-\pi \hbar c N_\perp I / 3 a k_B T} \right),
\label{eq:Lmax1}
\ee
where the magnon velocity $c$ is set by the curvature of the RKKY minimum, according to Eq.~\eqref{eq:long wavelength}. Equation \eqref{eq:Lmax1} can be further well approximated by
\be
L_{\rm max} \approx L_0 \exp\left( \frac{L_0}{a} N_\perp I\right),
\label{eq:Lmax2}
\ee
with the characteristic length scale $L_0 = \pi \hbar c /3k_B T$. Without trying to quantitatively estimate the magnon velocity $c$, we note that even for $T=1$ K, and $c=1$ m/s, by which we strongly underestimate $L_0$, we get $L_0\approx 0.01$ nm, and $L_{\rm max}$ is still exponentially large. This suggests that the long wavelength magnons can be safely neglected throughout this work.

\subsection{Peierls and Knight field energies}
\label{app:Tc2}
Once the nuclear order is established in an initially gapless electronic stripe, $\langle {\bf I}_i \rangle \neq 0$, the energy of the electrons is decreased by the opening of a partial gap. The change of this energy gain of the system per single flipped nuclear spin is called here the Peierls energy $\epsilon_P$. The opening of a gap also leads to a finite electron spin polarization $\langle {\bf S}(x_i) \rangle \neq 0$ locally collinear with the helical magnetic spin, which gives a Zeeman energy for the nuclear spin flip. We call this a Knight field energy $\epsilon_K$.

We estimate these two energies from the non-interacting electron model. This is partly justified by the fact that the gap opened at the Fermi energy suppresses the interaction effects, which predominantly arise from scattering at the Fermi energy. It is further illustrated by the result of Ref.~[\onlinecite{braunecker_nuclear_order_prb_09}], which found that the contribution to the RKKY exchange is greatly suppressed in the gapped subband. This can be seen as a suppression of the interaction-induced enhancement, back towards or even below the non-interacting value (depending on the value of the gap). Finally, the interaction effects are partially taken into account by the renormalization of $A \to A^*$, which we use also for the energies originating from the gapped subband.

Again within a given stripe, let us consider a basis of electron states $\Psi_{k \sigma}$ with longitudinal momentum $k$ and spin projection $\sigma$ along the helical axis with corresponding energies (we recall that we neglect the Zeeman energy) $\epsilon_{k \sigma} =\hbar^2 k^2/2m$. Within this basis, the Hamiltonian given in Eq.~\eqref{eq:Hhyp} has matrix elements 
\be
\Delta_{kk^\prime}^{\sigma \sigma^\prime}  \equiv \langle \Psi_{k\sigma} | H_{e-n} | \Psi_{k^\prime\sigma^\prime} \rangle = m\,\Delta \, \delta_{\sigma,-\sigma^\prime} \delta_{k+\sigma2k_F,k^\prime},
\label{eq:DeltaME}
\ee
where we denoted $\Delta=A S I$. To arrive at Eq.~\eqref{eq:DeltaME}, we replaced the nuclear spin operators by their expectation values in the presence of an established order according Eq.~\eqref{eq:ground state}. As follows from Eq.~\eqref{eq:DeltaME}, the electron states are pairwise coupled,
\be
H_k = 
\left( \begin{tabular}{cc}
$\epsilon_k$ & $m\Delta$\\
$m\Delta$ & $\epsilon_{k^\prime}$
\end{tabular} \right),
\label{eq:Hk}
\ee
where $k^\prime=k-2k_F$ for $k\in\langle 0, k_F\rangle$ and $k^\prime=k+2k_F$ for $k\in\langle -k_F, 0\rangle$. The eigenstates of Eq.~\eqref{eq:Hk}, denoted by $\Psi_{k\pm}$, correspond to eigenvalues
\be
\epsilon_\pm(k) = \frac{\epsilon_k + \epsilon_{k^\prime}}{2} \pm \sqrt{ \frac{(\epsilon_k - \epsilon_{k^\prime})^2}{4} +m^2\Delta^2 },
\ee 
so that $\epsilon_+>E_F>\epsilon_-$. 
At zero temperature, $\Psi_{k-}$ is occupied, and $\Psi_{k+}$ is empty. The change of the electronic band energy can be then obtained by summing the energies of the former throughout the band. The integral can be calculated analytically and we get in the leading order of small quantities ($\Delta/E_F$, and $\Delta/\Delta_a$)
\be
\epsilon_P= \frac{1}{\pi} \frac{m}{N_\perp} \frac{A^2}{\Delta_a} S^2 I \ln \left( \frac{2E_F}{m A S I} \right).
\label{eq:epsilon P}
\ee
The finite temperature effects are commented below.

We now turn to the calculation of the Knight field. The energy to decrease a nuclear spin by $\hbar$ in the presence of electron spin polarization $\langle {\bf S} \rangle \neq 0$ follows from Eq.~\eqref{eq:Hhyp} as
\be
\epsilon_K = A \langle {\bf S}(x_i) \cdot \frac{{\bf I}_i}{I} \rangle.
\ee
It relates by $\langle {\bf S} (x_i) \cdot {\bf I}_i \rangle =\langle P \rangle / N N_\perp$ to the operator of the total electronic polarization projected on the local helical order axis
\be
P = \sum_i {\bf S}(x_i) \cdot \frac{{\bf I}_i}{I}.
\ee
Since the latter is proportional to the electron-nuclear Hamiltonian itself, the matrix elements follow from Eq.~\eqref{eq:DeltaME} as
\be
P_{kk^\prime}^{\sigma \sigma^\prime} = S \delta_{\sigma,-\sigma^\prime} \delta_{k+\sigma2k_F,k^\prime}.
\ee 
The calculation then proceeds very similarly to the one for the Peierls energy, with the difference that we are now interested in the mean value of an operator $P_k$, which in the basis corresponding to Eq.~\eqref{eq:Hk} takes the form
\be
P_k=S\left( \begin{tabular}{cc}
0 & 1\\
1 & 0
\end{tabular} \right),
\label{eq:Pk}
\ee
so that
\be
\langle P \rangle = \sum_{k\leq k_F} \langle \Psi_{k-} | P | \Psi_{k-} \rangle.
\ee
A short calculation gives the Knight energy as
\be
\epsilon_K  = \frac{m}{2\pi N_\perp} \frac{A^2}{\Delta_a} S^2  \ln \left( \frac{2E_F}{m A S I} \right).
\label{eq:epsilon K}
\ee
This is a result obtained for non-interacting electrons. According to Ref.~[\onlinecite{braunecker_nuclear_order_prb_09}], we roughly estimate the interaction affects in the gapped sub-band by renormalizing the electron-nuclear coupling $A\to A^*$ using Eq.~\eqref{eq:A renormalization}, in both Eq.~\eqref{eq:epsilon P} and \eqref{eq:epsilon K} when plotting Fig.~\ref{fig:Tc}b.

Finally, we consider finite temperature effects. Both the Peierls and Knight field energies arise mainly from pairs of single electron states around the Fermi energy $\Psi_{k\pm}$, which are pushed away from each other by the energy $2m\Delta$. Their contribution to the Peierls gain, or the Knight field, are opposite. At a finite temperature, it is no more only the lower populated, but both, which leads to a suppression factor 
\be
p_-(T)-p_+(T)  \simeq 1 - \frac{2}{1+\exp(m \Delta/k_B T)}.
\label{eq:temperature suppression}
\ee
This factor should be inserted into the zero temperature expressions for $\epsilon_P$ and $\epsilon_K$. 
However, since we find that the critical temperature is much lower than $\Delta$, these effects are negligible and the factor in Eq.~\eqref{eq:temperature suppression} can be safely replaced by one.

\section{RKKY exchange of a gapless electron system, and the feedback effect}
\label{app:A2}
Let us now detail the RKKY exchange driving the ordering of the nuclear spins in the case of gapless stripes, where feedback effects are important. Because the tunneling has to be off resonant for the stripes to be gapless, we use the real space basis associated with the stripe index $n$, and neglect the RKKY exchange between different stripes (compared to the intrastripe exchange, the interstripe exchange is weakened by powers of the off-resonant interstripe tunneling). To calculate the RKKY exchange, we proceed along the lines of Ref.~[\onlinecite{braunecker_nuclear_order_prb_09}] and adopt a continuum model along a given stripe. The RKKY exchange is then calculated by evaluating the intrastripe spin susceptibility $\chi_{xx}^{\rm R}$ using 
\be
J_q = \frac{A^2a}{2 \hbar N_\perp^2}\chi_{xx}^{\rm R}(q,\omega\to0),
\ee
with $\chi_{xx}^{\rm R}(q,\omega)$ being the Fourier transform of the retarded susceptibility $\chi_{xx}^{\rm R}(x,t)$, defined by the following generalization of Eq.~\eqref{eq:FTdefinition} for continuous variables,
 \be
\chi^{\rm R}_{xx}(q,\omega)= \int dx\int dt\, e^{i(\omega t-q x)}\,\chi^{\rm R}_{xx}(x,t).
 \ee
The retarded spin susceptibility is defined as usual [\onlinecite{giamarchi_book}] through a Wick rotation of the imaginary time susceptibility

\begin{align}
& \chi_{xx}(x,\tau) \label{eq:chi}\\
 &= \frac{1}{4}e^{-i2xk_{F}}\langle T_\tau R_{n,\uparrow}^\dagger(x,\tau)L_{n,\downarrow}^\pdag(x,\tau)L_{n,\downarrow}^\dagger(0,0)R_{n,\uparrow}^\pdag(0,0)\rangle\nonumber\\
  &+ \frac{1}{4}e^{-i2xk_{F}}\langle T_\tau R_{n,\downarrow}^\dagger(x,\tau)L_{n,\uparrow}^\pdag(x,\tau)L_{n,\uparrow}^\dagger(0,0)R_{n,\downarrow}^\pdag(0,0)\rangle\nonumber\\
  &+\text{{\rm H.c.}}~,\nonumber\\
&= -\frac{1}{4(2\pi a)^2}e^{-i2xk_{F}}\nonumber\\
&\Bigl(\langle T_{\tau}e^{-i\sqrt{2}(\phi_{n,\rho}(x,\tau)-\theta_{n,s}(x,\tau)-\phi_{n,\rho}(0,0)+\theta_{n,s}(0,0))} \rangle\nonumber\\
&+\langle T_{\tau}e^{-i\sqrt{2}(\phi_{n,\rho}(x,\tau)+\theta_{n,s}(x,\tau)-\phi_{n,\rho}(0,0)-\theta_{n,s}(0,0))}\Bigr)+\text{{\rm H.c.}}\nonumber
\end{align}
Using these definitions, Ref.~[\onlinecite{braunecker_nuclear_order_prb_09}] found the RKKY exchange to be a function with a sharp minimum at $q=2 k_F$ with the minimal value
\be
J_{2k_F} (T)= -\frac{c(g)}{N_\perp^2}\frac{A^2}{\Delta_a} \left( \frac{\Delta_a}{k_B T} \right)^{2-2g} ,
\label{eq:J gapless}
\ee
where $k_B$ is the Boltzmann constant, the effective band width is $\Delta_a=\hbar v_F'(\Delta_t) /a$ with the effective Fermi velocity $v_F'(\Delta)$ given in Sec.~\ref{sec:IIIC}, the Luttinger liquid parameters define the exponent $g=(K_\rho+1/K_s)/2$, and where
\be
c(g)= \frac{\sin(\pi g)}{2}(2\pi)^{2g-4} \Gamma^2(1-g) \left| \frac{\Gamma(g/2)}{\Gamma(1-g/2)}\right |^2.
\label{eq:cg}
\ee
The width of the minimum is of order $\pi/\lambda_T$, with the thermal length $\lambda_T = \hbar v_F / k_B T$. 

\begin{figure}
\centering
\includegraphics[scale=0.35]{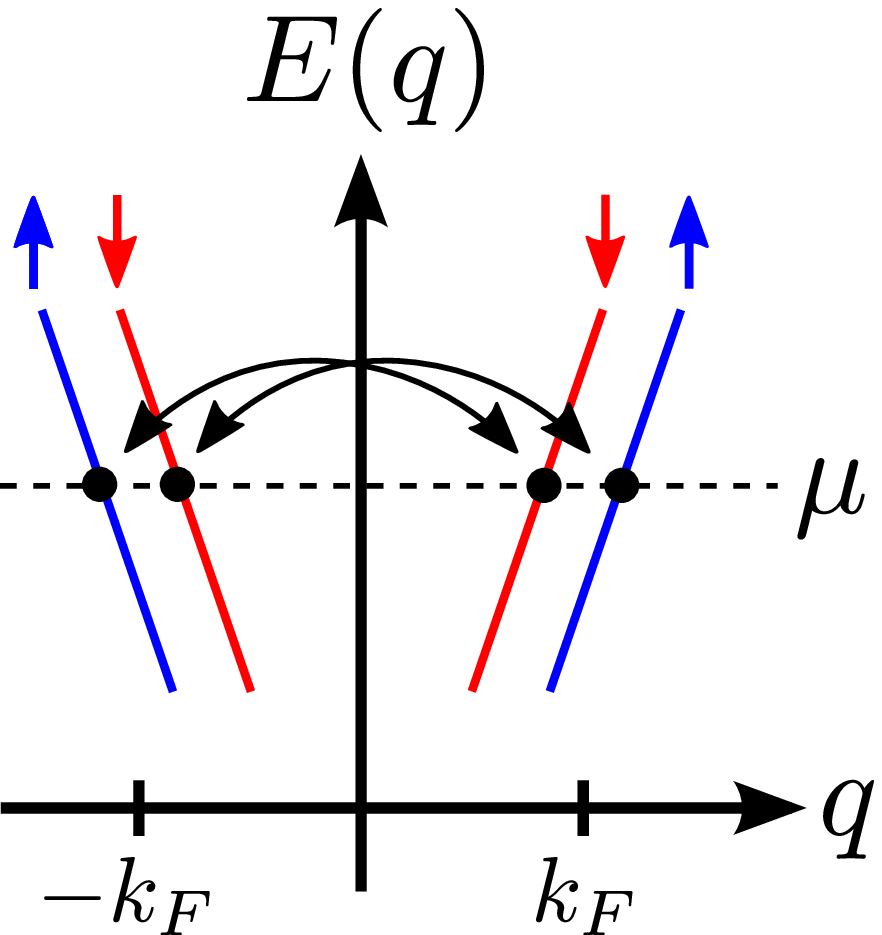}\quad \raisebox{1.53cm}{$\Rightarrow$} \quad\includegraphics[scale=0.35]{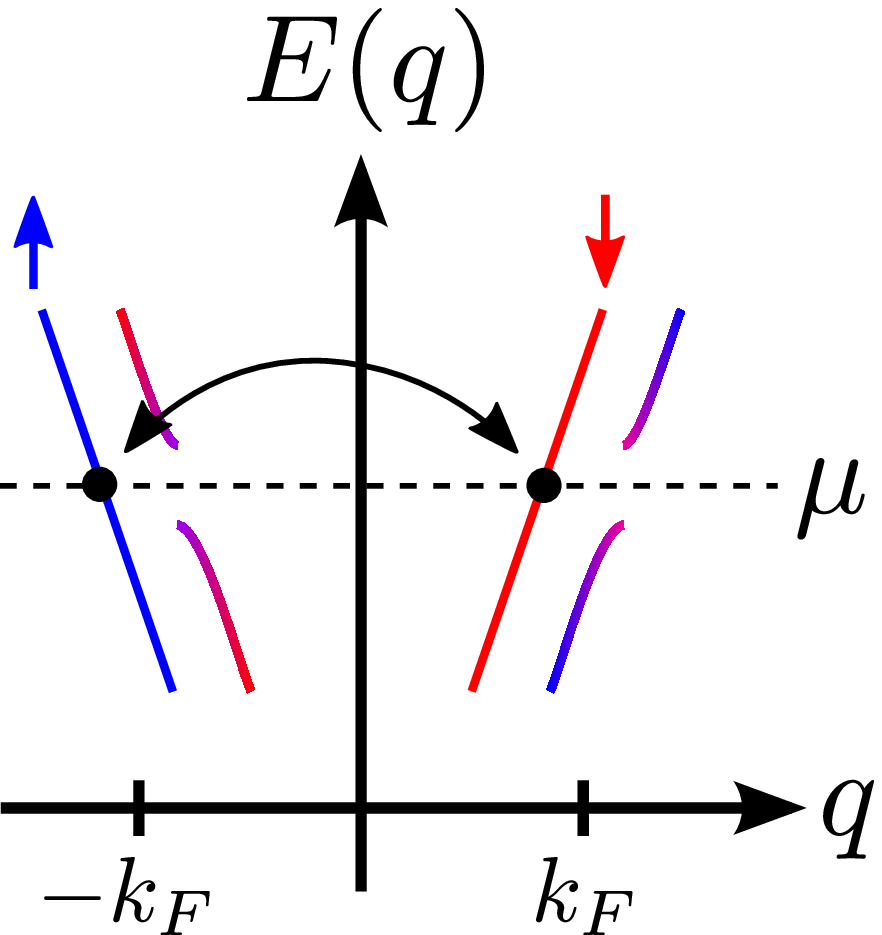}
\caption{(Color online) In a gapless stripe, the electrons mediate an RKKY interaction at momentum $q = 2k_{F}$ by spin-flip backscattering on the nuclear spins (black arrows - the degenerate spin up and spin down modes are offset for visibility). The resulting nuclear spin order gaps out parts of the spectrum in such a way that the remaining gapless modes provide a renormalized RKKY interaction, which stabilizes the order. The RKKY contribution from the gapped modes, on the other hand, is strongly suppressed.}
\label{fig:scatterings}
\end{figure}

The above result holds for an unperturbed (gapless) electronic system. With an order in the nuclear spins, the RKKY interaction is changed due to the opening of a partial gap in the electron system. Since this change of the RKKY interaction depends on the effects of the RKKY interaction itself (the nuclear order triggers the partial gap of the electrons), this is a highly non-trivial, non-linear feedback problem. First neglecting the effect of the feedback on the value of the RKKY exchange, Ref.~[\onlinecite{braunecker_nuclear_order_prb_09}] found that once the order is established, the electron subsystem can be thought of as being split into two subbands. One is gapless and mediates an RKKY interaction with the same functional form as in an unperturbed electron system, while the other subband is gapped. These subbands are depicted in Fig.~\ref{fig:scatterings}. The gapped subband turns out to give a negligibly small contribution to the RKKY exchange compared to the gapless part, but leads to additional energy gains. The feedback tends to enhance both the RKKY exchange of the gapless subband, and the gap of the other. The effects in the latter can be grasped by renormalization of the coupling constant 
\begin{equation}
A \to A^* = A (\xi/a)^{1-g},
\label{eq:A renormalization}
\end{equation}
 with the correlation length $\xi={\rm min} \{L,\lambda_T, a (\Delta_a/I A m)^{1/(2-g)}\}$. More important for our purposes are the feedback effects in the gapless subband, which strongly enhance the RKKY exchange, and consequently push the critical temperature to experimentally relevant values. This happens through a renormalization of the interaction exponent $g\to g''$, and the velocity $v_F\to v_F''$. The renormalization appears upon recalculating the RKKY exchange mediated by the gapless subband only [\onlinecite{braunecker_nuclear_order_prb_09}]. Using $u_i = v_{F}'(\Delta_t)/K_i$ (for $i=\rho,s$), we find

\begin{align}
 g'' =\frac{2K_\rho}{\sqrt{(1+K_\rho^2)(1+K_s^2)}}\label{eq:eff_g}
\end{align}
and 
\begin{align}
v_{F}''= v_F'(\Delta_t)\sqrt{\frac{1+K_\rho^2}{K_\rho^2+K_\rho^2K_s^2}}~.\label{eq:eff_vm}
\end{align}
The exponent $g''$, and the effective velocity $v_{F}''$ have already been discussed in Eqs.~(61) and (62) of reference [\onlinecite{braunecker_nuclear_order_prb_09}] (note that $\Gamma$ given in Eq.~(62) should be corrected to $4 \Gamma$), modulo a rescaling of the fields by a factor of $\sqrt{K_\rho/(K_\rho K_s+1)}$, as well as in Ref.~[\onlinecite{spectral_props_frac_ll}]. Because half of the low energy degrees of freedom are now gapped, the RKKY exchange furthermore acquires an additional factor of $1/2$ [\onlinecite{braunecker_nuclear_order_prb_09}]. Finally, we note that there is no feedback for a fully gapped stripe, since the gap prevents the helix to open a partial gap even if the nuclear order is established.

\section{RKKY interaction in a gapped stripe}\label{append:rkky_gapped}
As discussed in Sec.~\ref{sec:IIIB}, tunneling between the stripes can be brought to a simple form by Fourier transforming the fermionic Hamiltonian given in Eq.~\eqref{1d_y} along the direction perpendicular to the stripes. Unfortunately, the Fourier transformation renders the interaction given in Eq.~\eqref{eq:coulomb_ferm}, and consequently also the bosonized Hamiltonian given in Eq.~\eqref{eq:llham1}, highly off diagonal, such that we did not find an analytical expression for the RKKY exchange on a quantum Hall plateau. To nevertheless obtain an estimate for the order of magnitude of the RKKY exchange in a fully gapped phase, we analyze a different system, in which the spin conserving momentum-inverting interstripe hopping (giving rise to the quantum Hall effect) s replaced by a spin conserving momentum-inverting intrastripe scattering. While this state shares some similarities with the quantum Hall state, its main purpose here is to illustrate the suppression of the RKKY exchange in a fully gapped phase. Dropping the Klein factors, the intrastripe tunneling corresponds to a sine-Gordon term with the Hamiltonian density

\begin{align}
\mathcal{H}_{\rm intra} &= \sum_{n,\sigma} \frac{t}{\pi a}\label{eq:tunneling_ham_bos_intra}\cos(\sqrt{2}(\phi_{n,\rho}+\sigma \phi_{n,s}))~,
\end{align}
which is RG relevant for $K_\rho+K_s<4$. At the end of the RG flow, when the renormalized backscattering amplitude is of the order of the renormalized bandwidth, we expand the sine-Gordon term [\onlinecite{giamarchi_book}], and obtain

\begin{align}
 \mathcal{H} &=\sum_{n,\kappa=\rho,s} \frac{\hbar}{2\pi} \left(\frac{u_\kappa}{K_\kappa}(\partial_x \phi_{n,\kappa})^2+u_\kappa K_\kappa(\partial_x \theta_{n,\kappa})^2\right)\label{eq:llham3}\\
 &+ \sum_{n,\kappa=\rho,s} \frac{\hbar}{2\pi} \frac{\Delta_{t}^*{}^2}{\hbar^2 v_F}\phi_{n,\kappa}^2~.\nonumber
\end{align}
Here, $\Delta_{t}^*=2t(b^*)$ is the gap associated with $t$, where $2t(b^*)=\hbar v_F/a(b^*)$ defines the renormalized value of the running $t(b)$ in terms of the running short distance cutoff $a(b)$ the end of the RG flow, where $b=b^*$ [\onlinecite{braunecker_nuclear_order_prb_09,meng_nuclear_order_double_wire}]. We discriminate this model from the original QHE model by the notation, introducing $\Delta_t^*$ as an analog of the QHE bulk gap $\Delta_t$.

\begin{figure}
\centering
\includegraphics[scale=0.65]{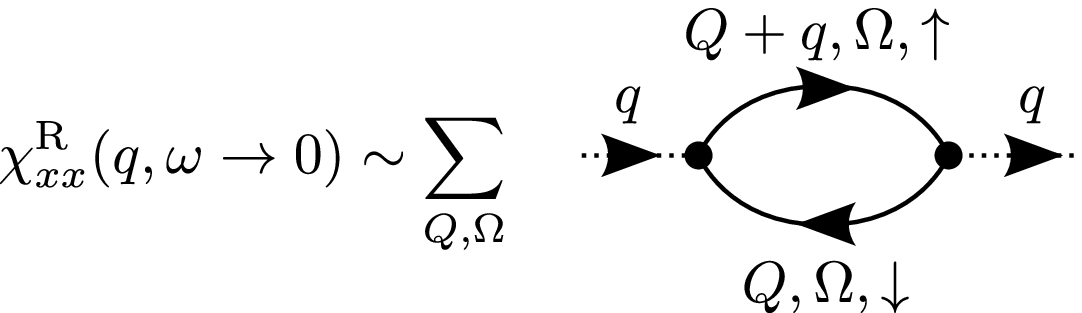}
\caption{The particle hole bubble determining the spin susceptibility $\chi^R_{xx}(q,\omega\to0)$.}
\label{fig:susc_bubble}
\end{figure} 

We now proceed to evaluate the RKKY exchange in a stripe fully gapped by intrastripe backscattering. To this end, we first calculate the spin susceptibility, defined in Eq.~\eqref{eq:chi}, using the Hamiltonian density given in Eq.~\eqref{eq:llham3}. At frequencies and momenta smaller than the gap $\Delta_{t}^*$, the fields $\phi_{n,\rho}$ and $\phi_{n,s}$ are both pinned to constant values. At frequencies and momenta higher than the gap, on the contrary, the fields can overcome the pinning, and the theory essentially recovers its gapless form (since at these frequencies and momenta, $\Delta_{t}^*$ is negligible compared to the kinetic terms). This scale dependence is reflected in the form of the correlator invoking the gapped fields $\phi_{n,\rho}$,
\begin{align}
\langle(\phi_{n,\rho}&(x,\tau)-\phi_{n,\rho}(0,0))^2\rangle\nonumber\\
&=\frac{1}{\hbar\beta L}\sum_{q,\omega_n}\frac{2\pi u_\rho K_\rho [1-\cos(qx-\omega_n\tau)]}{\omega_n^2+u_\rho^2q^2+\Delta^*_{t}{}^{2}/K_\rho \hbar^2}.\label{eq:prop_gap}
\end{align}
Here, $\beta^{-1}=k_BT$ is the inverse temperature and $L$ the length of the stripe. At temperatures smaller than the gap $\Delta_{t}^*$, and introducing a UV cutoff $\hbar v_F/a\gg\Delta_{t}^*$, we can approximate this correlator by its zero temperature expression
\begin{align}
&\frac{1}{K_\rho}\langle(\phi_{n,\rho}(x,\tau)-\phi_{n,\rho}(0,0))^2\rangle\nonumber\\
&= K_0\left(\frac{a \Delta_{t}^* \sqrt{K_\rho}}{\hbar v_{F}}\right)\\
& - K_0\left(\frac{\Delta_{t}^* \sqrt{K_\rho}\sqrt{x^2+(u_{\rho}|\tau|+a)^2}}{\hbar v_{F}}\right)\nonumber~,
\end{align}
where $K_0(x)$ is a modified Bessel function of the second kind. The latter behaves as $K_0(x)\approx-\ln(x)$ for $x\ll1$, and is exponentially suppressed at $x\gg1$.

The dependence of the propagators on the energy scale $\Delta_{t}^*$ is inherited by the spin susceptibility. To illustrate this, we recall that the spin susceptibility given in Eq.~\eqref{eq:chi} corresponds to the exchange of particle hole pairs (``bubble'') as depicted in Fig.~\ref{fig:susc_bubble}. The internal sum over frequency and momenta can be divided into frequencies and momenta smaller than the gap, $v_F |Q|, v_F |q+Q|, |\Omega| < \Delta_{t}^*$, and larger than the gap. As can be inferred from Eq.~\eqref{eq:prop_gap}, the gap regularizes the contribution stemming from small momenta and frequencies, which would diverge in the absence of a gap and at zero temperature, while the contribution stemming from large frequencies and momenta is basically unaffected by the gap. Following references [\onlinecite{starykh_gapped_99,braunecker_prb_12,meng_spin_suscept}], we evaluate the spin susceptibility first in real space and imaginary time, which yields

\begin{align}
&\chi_{xx}(x,\tau) \approx \begin{cases}\chi_{xx,<}(x,\tau),~\sqrt{x^2+u_{\rho}^2\tau^2}\ll\frac{\hbar v_F}{\Delta_{t}^*\sqrt{K_\rho}}\\\chi_{xx,>}(x,\tau),~\sqrt{x^2+u_{\rho}^2\tau^2}\gg\frac{\hbar v_F}{\Delta_{t}^*\sqrt{K_\rho}}~,\end{cases}\label{eq:cases}
\end{align}
with
\begin{align}
&\chi_{xx,<}(x,\tau)\approx -\frac{1}{2(2\pi a)^2}e^{-i2xk_{F}}\left(\frac{a }{\sqrt{x^2+(u_\rho|\tau|+a)^2}}\right)^{K_\rho}\nonumber\\
&\times\left(\frac{a}{\sqrt{x^2+(u_s|\tau|+a)^2}}\right)^{1/K_s}+\text{{\rm H.c.}}~,\label{eq:gapless_sus}
\end{align}
and
\begin{align}
&\chi_{xx,>}(x,\tau)\approx -\frac{1}{2(2\pi a)^2}e^{-i2xk_{F}}\,\left(\frac{a\Delta_{t}^*\sqrt{K_\rho}}{\hbar v_F}\right)^{K_\rho}\\
&\times\left(\frac{a}{\sqrt{x^2+(u_s|\tau|+a)^2}}\right)^{1/K_s}\,e^{-\mathcal{C}\frac{\Delta_{t}^*}{\hbar v_F}\sqrt{x^2+u_s^2\tau^2}}+\text{{\rm H.c.}}~,\nonumber
\end{align}
where $\mathcal{C}$ is a constant of order one. As expected, Eq.~\eqref{eq:cases} shows that at length and time scale larger than the correlation length associated with the gap $\Delta_{t}^*$, the spin susceptibility is exponentially suppressed, while it approaches its gapless form at small length and times scales.

Neglecting the difference between $u_\rho$ and $u_s$, setting $\mathcal{C}\to1$, performing a (continuous) Fourier transformation to momentum and Matsubara frequencies as well as the analytic continuation $i\omega_n\to\omega+i0^+$, and taking the static limit $\omega\to 0$ finally shows that the largest contribution to the static spin susceptibility stems from small distance and time scales. It is approximately given by  
\begin{align}
&\chi_{xx,<}^{\rm R}(q\approx2k_F,\omega\to 0)\\
&\approx \frac{-1}{4\pi v_F}\frac{1}{2-K_\rho-1/K_s}\left(\left(\frac{\hbar v_F\sqrt{K_\rho}}{\Delta_{t}^* a}\right)^{2-K_\rho-1/K_s}-1\right)\nonumber~.
\end{align}
Taking the limit $K_\rho$, $K_s\to1$ on this expression, we recover the logarithmic dependence of $\chi_{xx}^{\rm R}$ on $\Delta_{t}^*$ that has been obtained in the non-interacting case using a fermionic calculation [\onlinecite{klinovaja_sc,braunecker_sc}]. Due to its exponential suppression, the contribution from $\chi_{xx,>}^{\rm R}$ essentially derives from fluctuations close to the length scale associated with the gap, and consequently has a similar dependence on $(\Delta_{t}^* a/\hbar v_F)^{K_\rho+1/K_s-2}$ for interacting systems, albeit with a smaller prefactor. This contribution reduces to a constant, independent of $\Delta_{t}^* a/\hbar v_F$, in the non-interacting limit. The largest contribution to the total spin susceptibility thus stems from $\chi_{xx,<}^{\rm R}$, and derives from fluctuations on length scales of the order of, or smaller than, $\hbar v_F/\Delta_{t}^*$. Importantly, the spin susceptibility is independent of temperature, provided the latter is smaller than the gap (simply speaking, the zero temperature Luttinger liquid divergence obtained for a gapless stripe is cut off by the maximum of temperature and gap). Therefore, the critical temperature defined via Eq.~\eqref{eq:Tc final} scales as $1/N_\perp$, and is largely suppressed compared to a gapless stripe. Based on these considerations, we approximate the RKKY exchange as

\begin{align}
& J_{2k_F} \approx -\frac{1}{4\pi}\frac{A^2a}{\hbar v_F N_\perp^2}\\
 &\times\frac{1}{2-K_\rho-1/K_s}\left(\left(\frac{\hbar v_F\sqrt{K_\rho}}{\Delta_{t}^* a}\right)^{2-K_\rho-1/K_s}-1\right)~.\nonumber
\end{align}



\begin{thebibliography}{99}
\bibitem{footnote_pol}
Spin polarized electrons cannot mediate a strong RKKY interaction, the resonant enhancement of which requires electron spin-flip backscattering. We thus expect that in the spin polarized regime, dynamical polarization effects are more important than a presumably negligible order of the nuclear system in thermodynamic equilibrium. This is an entirely different situation than the one we consider in this article, and requires a separate analysis which we postpone to a future work.
\bibitem{footnote_dens}
The Hamiltonian density $\mathcal{H}(x)$ relates to the Hamiltonian $H$ by $H = \int {\rm d} x\, \mathcal{H}(x)$, a notation we will use similarly also in further.
\bibitem{footnote1}
To simplify the presentation, we will approximate this value by replacing the Brillouin function by the Langevin function $B_I(x) \to L(x)=\coth(x)-1/x$, and Taylor expanding the latter, $L(x) \approx x/3$, resulting in $k_B T_c = (I/3) \epsilon(m)/m$, which we use to get Eq.~\eqref{eq:Tc final}.
\end{thebibliography}
\end{document}